\documentclass[english,aps,pra,twocolumn,superscriptaddress,SISSA]{revtex4}
\usepackage[T1]{fontenc}
\usepackage[latin1]{inputenc}
\usepackage{graphicx}
\usepackage{amssymb}
\usepackage{sublabel}
\usepackage{color}
\usepackage{textcomp}
\usepackage{subfigure}
\usepackage{amsmath}
\makeatletter

\usepackage{babel}
\makeatother
\begin{document}

\title{Cortical free association dynamics: distinct phases of a latching network}

\author{Eleonora Russo}
\affiliation{SISSA, Cognitive Neuroscience, via Bonomea 265, 34136 Trieste, Italy}
\homepage{http://www.sissa.it/~ale/limbo.html}
\email{russo@sissa.it}
\author{Alessandro Treves}
\affiliation{SISSA, Cognitive Neuroscience, via Bonomea 265, 34136 Trieste, Italy}


\date{\today}

\begin{abstract}
A Potts associative memory network has been proposed as a simplified model of macroscopic cortical dynamics, in which each Potts unit stands for a patch of cortex, which can be activated in one of $S$ local attractor states. The internal neuronal dynamics of the patch is not described by the model, rather it is subsumed into an effective description in terms of graded Potts units, with adaptation effects both specific to each attractor state and generic to the patch. If each unit, or patch, receives effective (tensor) connections from $C$ other units, the network has been shown to be able to store a large number $p$ of global patterns, or network attractors, each with a fraction $a$ of the units active, where the critical load $p_c$ scales roughly like $p_c\approx C S^2/a \ln (1/a)$ (if the patterns are randomly correlated). Interestingly, after retrieving an externally cued attractor, the network can continue jumping, or {\em latching}, from attractor to attractor, driven by adaptation effects. The occurrence and duration of latching dynamics is found through simulations to depend critically on the strength of local attractor states, expressed in the Potts model by a parameter $w$. Here we describe with simulations and then analytically the boundaries between distinct phases of no latching, of transient and sustained latching, deriving a phase diagram in the plane $w-T$, where $T$ parametrizes thermal noise effects. Implications for real cortical dynamics are briefly reviewed in the conclusions. 
\end{abstract}

\maketitle

\section{Introduction}

The Hopfield \cite{Hop82} model and its subsequent analysis with statistical physics techniques \cite{Ami+87} have provided an important quantitative theoretical framework to approach the operation of associative memory networks in the brain. After many years of modeling developments and gradual familiarization by the neuroscience community \cite{Ami95,Rol+98}, the notion of attractor states forcefully put forward by the Hopfield model has guided experimental investigations in the rat hippocampus \cite{Wil+05,Leu+05} and in the primate inferotemporal cortex \cite{Ami+97,Akr+09}, suggesting analyses of the time course of attractor dynamics which have revealed a striking rhythmically pulsed or "flickering" regime \cite{Jez+11}.

\begin{figure}[ht]
\begin{center} 
\includegraphics[width=.45\textwidth]{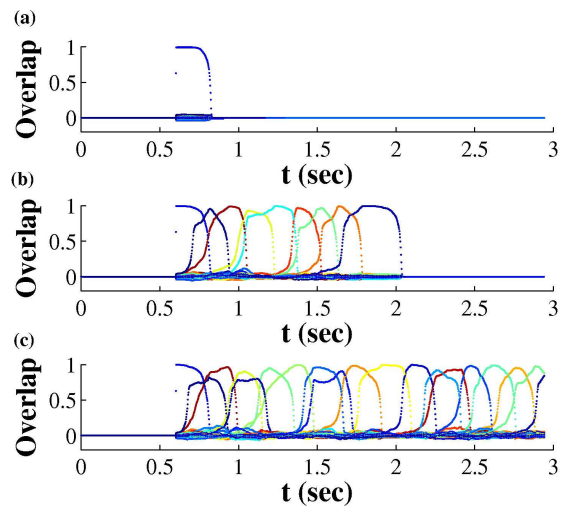}
\caption{(Color online) Examples of latching dynamics in the Potts network. The correlation (overlap) of the network state with each of a number of stored activity patterns is indicated in a different color. The retrieval of a memory can be followed by the inactive state (a) or by latching dynamics, that either die out after a few steps (b) or self-sustain indefinitely (c), depending on the parameters of the network.}
\label{fig:lungo}
\end{center} 
\end{figure}

Attractor dynamics has been proposed to be relevant to global cortical function \cite{Bra78,Bra+91}, also in situations in which the cortex actually produces behavior, e.g. speech, confabulation, or a drawing, as opposed to just responding to a stimulus from outside \cite{Pul05,Mit+08,Rus00,Bur+96,Shm+06,Sos+07,Abe04}. It appears then of particular interest to go, within a modeling framework, beyond the standard cued retrieval operation, to protracted cortical dynamics including attraction to, and escape from, a sequence of metastable states, or attractor "ruins" \cite{Kan+03,Tsu92}. Firing frequency adaptation is a simple and salient feature of neuronal dynamics in the cortex, which results in the unavoidable metastability of any of its attractor states. The study of a full neuronal network model of an extended "cortex" with firing frequency adaptation is complicated, however, by the necessity to treat simultaneously local and global dynamics \cite{O'K+92a}. 
The Potts associative memory network model \cite{Kan88,Bol+93} can be interpreted as a global cortical model, in which local dynamics are subsumed into an effective description \cite{Tre05,Akr+11}. Its storage and retrieval capacity (without adaptation) can be studied analytically \cite{Kro+05}. With adaptation, it shows latching behavior, i.e. a sequence of jumps from one attractor to another (Fig.~\ref{fig:lungo}), with non-trivial transition statistics, neither random nor deterministic \cite{Rus+11, Cos+09}.
The aim of the present study is to characterize the conditions for latching dynamics to occur, and to self-sustain indefinitely. The model is defined in the next section, and the various latching regimes are described in Sec.~\ref{sec:latch}. Then in Sec.~\ref{sec:analytical} an analytical approach is developed, that yields a phase diagram for the model. Its implications for real cortical dynamics are reviewed in the last section.

\section{The model}\label{sec:model}

The network considered here is the same as studied previously~\cite{Tre05} and we refer to that paper and to Ref.~\cite{Akr+11} for a discussion of the correspondence with real cortical networks. Here we dwell on the mathematical description of the model, based on the hypothesis that, abstracted from observed features of cortical representations~\cite{Lac+09}, it may be relevant to a better understanding of cortical dynamics, over time scales of up to a few seconds.

\subsection{Potts units}\label{sec:pottsunits}

Let us consider a network comprised of many subnetworks interacting with each other with long range connections. 
Each of these small networks has its own attractors, with which it tends spontaneously to align during the dynamics. 
So, while in a brief transient condition the activation state of each small network may be partially overlapping with several of its attractors, during a protracted quasi-stable memory retrieval condition its alignment with one of the attractors tends to be near complete.

\begin{figure}[ht]
\begin{center}
\includegraphics[width=.4\textwidth]{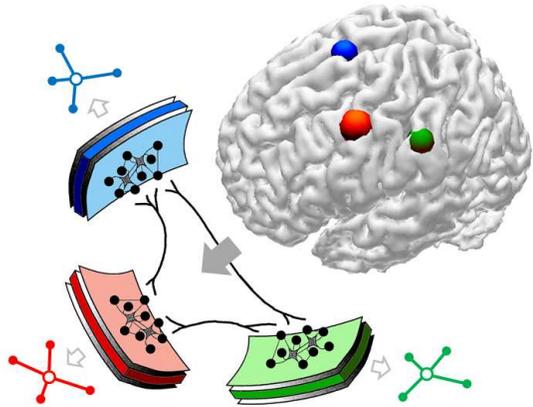}
\caption{(Color online) Conceptual derivation of the Potts network, obtained as a simplification of a two-tier associative memory network by reducing its local dynamics to single Potts units, representative of the states of patches of cortex.}
\label{fig:net}
\end{center}
\end{figure}

To summarily describe these local dynamics we represent the small subnetworks with Potts units, each of which can be active, to a variable degree from $0$ to $1$, in a number $S$ of distinct active states, representing local attractors, and in one inactive state, when the local activity pattern has no correlation with any of the attractors (Fig.~\ref{fig:net}). At any time, the sum of the activity levels of unit $i$, $\sigma_i^k$, $k\in\left[0,S\right]$, ($k=0$ denotes the inactive state) is kept fixed to $1$, $\sum_{k=0}^S\sigma_i^k=1$.

The model thus reduces to an autoassociative network of $N$ Potts units bound by long range connections. The connections, stored in the weight matrix of the network, are fixed, with the standard assumption \cite{Hop82} that they reflect an earlier learning phase. 
By setting the weights, one defines the global attractors of the system. These are configurations of states of Potts units, that represent combinations of local attractors in each subnetwork, and in turn serve as attractors for the global dynamics of the Potts units. In particular we take the network to have stored as global attractors $p$ activity patterns $\xi^\mu$ (see Sec.~\ref{sec:corrpatt}), with $\mu =1,\dots,p $. The strength $J_{ij}^{kl}$ of the connection between state $k$ of unit $i$ and state $l$ of unit $j$ reads \cite{Bol+93,Kro+05}
\begin{equation}
J_{ij}^{kl}=\frac{c_{ij}}{C a(1-\frac{a}{S})}\sum_{\mu=1}^{p}(\delta_{\xi_{i}^{\mu}k}-\frac{a}{S})(\delta_{\xi_{j}^{\mu}l}-\frac{a}{S})(1-\delta_{k0})(1-\delta_{l0})
\end{equation}
where $C$ is the average number of units connected with a given unit (denoted as $c_M$ in \cite{Tre05} to indicate connectivity between Modules) and $a$ is the {\em sparsity} parameter, i.e. the fraction of units in an active state in each of the stored patterns \cite{Tre05}. The connection matrix element $c_{ij}$ is equal to $1$ if the presynaptic unit (module) $j$ is connected to the postsynaptic unit (module) $i$, and $0$ otherwise. The connection tensor
$J_{ij}^{kl}$ can be written more succinctly by dropping the last two factors and intending it to act only between active states, given that
the inactive state of each unit, $\sigma_i^0$, is not an independent variable because of the constraint $\sum_{k=0}^S\sigma_i^k=1$. 

A simple update rule can be assigned to the network in discrete time, by randomly selecting each unit at a time, evaluating the {\em local field} in the direction of its different states $h_i^k=\sum_j J_{ij}^{kl} \sigma_j^l$ (plus any input from outside the Potts network), and aligning the unit in the direction $k^*$ of the strongest field if this surpasses a threshold $U$, $\sigma^{k^*}=1$, and setting it in the inactive state, $\sigma^0=1$, otherwise. The thermodynamics of such a model, for $c_{ij}\equiv c_{ji}$, has been analysed by \cite{Kro+05}, deriving it from the Hamiltonian
\begin{equation}
H=-{1\over 2}\sum_{i,j\ne i}^N\sum_{k,l=1}^S J_{ij}^{kl}\sigma_i^k\sigma_j^l +U\sum_i^N\sum_{k=1}^S \sigma_i^k
\end{equation}
which can be thought to act under the influence of noise, modeled by introducing a non-zero temperature $T\equiv\beta^{-1}$.
The network stabilizes into one of several possible asymptotic states, the global attractors, which may include clusters of configurations around the quiescent state $\{\sigma_i^0\equiv 1\}$ as well as around the memory attractors $\{\sigma_i^{\xi^\mu}=1\}$ (all other states of a unit in the low noise regime being at $\sigma_i\simeq 0$). For $T=\beta^{-1}=0$ (no thermal noise at all) the asymptotic states reduce to a single configuration, and no further updating occurs. 

We aim here for a more realistic description of local cortical dynamics, which occurs in continuous time and includes a role for neuronal fatigue, to see the resulting effects on global dynamics.


\subsection{Adaptive dynamics}
\label{sec:Adaptivedynamics}

Transitions among attractors may be induced by a variety of mechanisms that either weaken temporarily the strength of the current attractor or produce an enhanced cue in the direction of another one \cite{Som+86,Hor+89,Her+93,Rab+01,Met+01,Tim+02,Pau+06,Gro07}.


The stability of the stored memories is weakened, in the model considered here, by the adaptive thresholds introduced to model neuronal fatigue and slow inhibition. Two types of time-varying thresholds are included: an additional contribution to the constant baseline threshold $U$, affecting all active states of a Potts unit, $\theta^0_i(t) $, with a time constant $\tau_3$ of hundreds of {\em ms} for its dynamics, intended to model various forms of slow, delayed inhibition within a cortical patch \cite{Tam+03,Koh+10,Fin+11}, and a specific threshold $\theta^k_i(t)$, affecting only Potts state $k$, with a time constant $\tau_2$ of tens of {\em ms} for its dynamics, intended to model resource depletion of the neurons (fatigue proper) and of the synapses (short-term depression \cite{Tso+97}) active in that state. With these modifications, after the stabilization into a global pattern, all the active Potts units are increasingly affected by fatigue and, with two characteristic time constants, tend to inactivate or to change active state, eventually changing the activity configuration of the network. Thereby the system is pushed away from the basin of attraction it had settled in and, sometimes, towards that of a new correlated memory attractor.

The activity of each unit is determined by a set of specific input variables $r_i^k(t)$, which rapidly integrate, with time constant $\tau_1$ (meant to be in the {\em ms} range), the {\em local fields}, i.e. the summed influence $h_i^k(t)$ of presynaptic units, subject to the dynamic thresholds $\theta_i^k(t)$, while the non-specific thresholds $\theta_i^0(t)$ and $U$ can be thought of as inputs to the inactive state.
In detail, the activation of the unit $i$ is set at any time $t$ by the non-linearity
\begin{equation}
\sigma_i^k=\frac{\exp(\beta r_i^k)}{\sum_{l=1}^S\exp(\beta r_i^l)+\exp[\beta (\theta_i^0+U)]}\label{eq:sigmaofr}
\end{equation}
for each active state $k$ and
\begin{equation}
\sigma_i^0=\frac{\exp[\beta (\theta_i^0+U)]}{\sum_{l=1}^S\exp(\beta r_i^l)+\exp[\beta(\theta_i^0+U)]}
\end{equation}
for the null state, with $\beta$ an inverse temperature parametrizing thermal noise, as in the Hopfield model~\cite{Hop82}. Note that $T\equiv \beta^{-1} > 0$ implies a smooth transfer function for the Potts units, but no stochasticity in their dynamics. The temperature in other words represents here thermal noise in the local network underlying each Potts unit, but the global network is noiseless. See \cite{Abd+12} for the analysis of a related model but with different, noisy dynamics.

The inputs are linearly integrated
\begin{equation}
\tau_1{dr_i^k(t)\over dt}=h_i^k(t)-\theta_i^k(t)-r_i^k(t).
\end{equation}
with 
\begin{equation}
h_i^k=\sum_{j\ne i}^N\sum_{l=1}^SJ_{ij}^{kl}\sigma_j^l+w(\sigma_i^k-\frac{1}{S}\sum_{l=1}^S\sigma_i^l)
\end{equation}
where it is important to note that we have added a self-reinforcement term with coefficient $w$, to model the nonlinear convergence towards the more active state, induced by the local patch dynamics not explicitly represented in the Potts model. The $w$ term is then a key difference between the original Potts autoassociative network model, as studied by Kanter \cite{Kan88}, and our version, intended as an effective model of a two-tier network. The parameter $w$ provides a positive feedback, that makes the convergence towards a local state a rapid self-regenerative process; it can be argued to numerically reflect the ratio of local to external afferent inputs, to each neuron in the two-tier network \cite{Rus+12}. We shall call it the {\em local feedback} term.

The adaptive thresholds follow, with slower dynamics, the mean activity of the unit
\begin{equation}
\noindent
\tau_2{d\theta_i^k(t)\over dt}=\sigma_i^k(t)-\theta_i^k(t)
\end{equation}
and 
\begin{equation} \label{eq:eqb3}
\tau_3{d\theta_i^0(t)\over dt}=\sum_{k=1}^S\sigma_i^k(t)-\theta_i^0(t).
\end{equation}

These equations completely define the dynamics of the network. 

Note that in the limit $\tau_2,\tau_3\to\infty $ the model sheds the adaptive character of its dynamics, and genuine attractor states are indefinitely stable.

Stability is also maintained when attractors are exceedingly "deep", that is, if the local feedback $w$ is strong enough to overcome the effects of the rising thresholds. Then the thresholds stabilize at the asymptotic values $\theta_i^{k} \to\bar{\sigma}_i^{k}$ and $\theta_i^{0} \to\sum_k\bar{\sigma}_i^{k}$, and the input variables $\{r\}$'s satisfy the asymptotic system $r_i^{k}=h_i^{k}-\theta_i^{k}$, where both the $\{h\}$'s and $\{\theta\}$'s are functions of the $\{\bar{\sigma}\}$'s, hence of the $\{r\}$'s through Eqs.~\ref{eq:sigmaofr}. Away from those limit cases, adaptation tends to destabilize memory attractors.

The one state that is guaranteed to remain stable (provided the constant threshold $U$ takes a positive value) is the "global null state", where all active states take low values, vanishing exponentially with $\beta$, and $\sigma_i^0\simeq 1$ (the exact values depend on $\beta$ and $U$).

\subsection{The slowly adaptive regime}\label{sec:slowregime}

The presence of adaptive thresholds makes an analytical approach to the study of the dynamics quite difficult, in general. Were it not for the adaptive thresholds, and assuming symmetric connectivity, $c_{ij}\equiv c_{ji}$, one would consider the energy function
\begin{equation}
E'=-\frac{1}{2}\sum_{i}^N\sum_{k=1}^S h_i^k \sigma_i^k+U\sum_i^N\sum_{k=1}^S \sigma_i^k.
\end{equation}
Due to the moving thresholds, however, no energy or free-energy can be defined, to also describe how the thresholds themselves depend on the activity.
Nevertheless, we are not so interested in the model in general, but rather specifically in the regime in which neuronal dynamics is faster than threshold dynamics, both intended at the local population level, i.e. when $\tau_3 > \tau_2 \gg \tau_1$ (the {\em slowly adaptive regime}). 

In the slowly adaptive regime the attractors, which were asymptotically stable without adaptation, gradually become unstable. By going to the limit in which $\tau_1$ is much shorter than the slow $\tau_2$ and $\tau_3$ times, one can conceptualize dynamics as occurring over completely separate time scales. 

In a time of order $\tau_1$ the $\{\sigma\}$ configuration relaxes towards an attractive metastable state, determined by nearly constant values for the thresholds, following a continuous time version of the discrete-time updating dynamics mentioned in Sec.~\ref{sec:pottsunits}. Having found a metastable state, an "attractor ruin" in the language of Tsuda \cite{Tsu92}, the $\{\sigma\}$'s can be rigidly entrapped there, especially if thermal noise if very low, $\beta \gg 1$. At this point neural dynamics becomes almost stationary, while thresholds continue to evolve, albeit slowly. On a time scale $\tau_2$, and with nearly constant $\{\sigma\}$'s, the thresholds change linearly until they destabilize the $\{\sigma\}$'s again. In this joint limit $\tau_2, \tau_3 \gg \tau_1$ and $\beta \gg 1$, then, and again assuming symmetric connectivity, one can describe separately the two dynamic modes, each governed by a distinct functional.

The rapid transient dynamics towards an attractor minimize the classical Energy 

\begin{widetext}
\begin{equation}
\begin{split}
E(\{\sigma (t)\}|\{\theta\}) = &-{1\over 2}\sum_{i,j\ne i}^N\sum_{k,l=1}^S J_{ij}^{kl}\sigma_i^k\sigma_j^l - {w\over 2}\sum_i^N\left[\sum_{k=1}^S(\sigma_i^k)^2 - {1\over S} (\sum_{k=1}^S\sigma_i^k)^2\right]+\label{eq:energy}\\
&+ \sum_i^N\sum_{k=1}^S \left\{ (U+\theta_i^0+\theta_i^k) \sigma_i^k +\int_0^{\sigma_i^k} \left[r_i^k (\{\sigma_i^1,\dots, \varrho^k,\dots,\sigma_i^S\})-U-\theta_i^0\right]d\varrho^k\right\}
\end{split}
\end{equation}
\end{widetext}
where we have just added the extra local feedback term (the second in the first row), and the thresholds appear as parameters, in the second row, while the second term in the second row follows the formulation proposed for quasi-binary graded response units by Hopfield \cite{Hop84}. This last term can be rewritten, considering that $r_i^k=U+\theta_i^0+(1/\beta)\ln (\sigma_i^k/\sigma_i^0)$ and carrying out the integrals, as
\begin{eqnarray}
\begin{split}
& E(\{\sigma (t)\}|\{\theta\})|_{\rm graded \; response \; term} =\\
&={1\over \beta}\sum_i^N\sum_{k=1}^S \left\{  \sigma_i^k \ln {\sigma_i^k\over \sigma_i^k+\sigma_i^0}+\sigma_i^0 \ln {\sigma_i^0\over \sigma_i^k+\sigma_i^0}\right\}\label{eq:energy2}
\end{split}
\end{eqnarray}
which clarifies its nature as an "entropy" term.

The slow adaptation dynamics minimize instead the Adaptation function
\begin{eqnarray}
\begin{split}
A(\{\theta (t)\}|\{\sigma\})&={1\over 2}\sum_{i}^N\sum_{k=1}^S \left[(\theta_i^k)^2-2\theta_i^k\sigma_i^k\right] + \\
&+ {1\over 2}\sum_{i}^N\left[(\theta_i^0)^2-2\theta_i^0\sum_{k=1}^S\sigma_i^k\right],\label{eq:adapt}
\end{split}
\end{eqnarray}
where now the activation variables appear as parameters.

\begin{figure}[ht]
\begin{center}
\includegraphics[width=.5\textwidth]{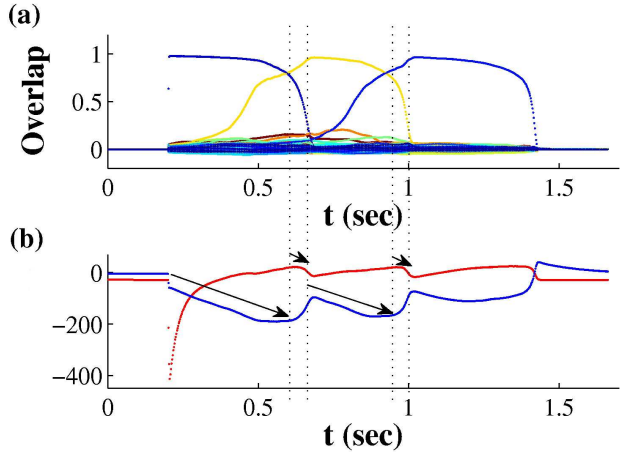}
\caption{(Color online) (a) Latching transitions; (b) corresponding value of the Energy (red curve) and Adaptation function (blue curve). During a transition, on a fast time scale, the Energy decrease drives the system to a new attractor state. Then, on a slower time scale, Adaptation destabilizes the attractor, pushing the network towards a new transition. As in Fig.~\ref{fig:lungo}, time is expressed in {\em sec}, assuming $\tau_1=10~msec$.}
\label{fig:energia}
\end{center} 
\end{figure}

Note that if we assume that over certain temporal intervals the thresholds $\{\theta\} $ are roughly constant,
we have
\begin{eqnarray}
{dE\over dt}&=&\sum_{i}^N\sum_{k,l}^S {\delta E\over\delta \sigma_i^k}{\delta \sigma_i^k \over \delta r_i^l}{dr_i^l\over dt}=- \tau_1 \sum_{i}^N\sum_{k,l}^S {dr_i^k\over dt}{\delta \sigma_i^k \over \delta r_i^l}{dr_i^l\over dt} = \nonumber \\
 &=&{} - \beta\tau_1 \sum_{i}^N\sum_{k,l}^S {dr_i^k\over dt}\sigma_i^k(\delta_{kl}-\sigma_i^l){dr_i^l\over dt}\le 0.\label{eq:denergydt}
\end{eqnarray}
because the matrix $\sigma_i^k(\delta_{kl}-\sigma_i^l)$, which has determinant $\Delta=\prod_{k=0}^S \sigma_i^k$ and trace ${\rm Tr}=\sum_{k=1}^S\sigma_i^k(1-\sigma_i^k)$ can be seen by induction to be positive definite; whereas if we assume that over other temporal intervals the attractor variables $\{\sigma\}$'s are roughly constant, we have instead
\begin{eqnarray}
\begin{split}
{dA\over dt}&=\sum_{i}^N\left[\sum_{k}^S {\delta A\over\delta \theta_i^k}{d\theta_i^k\over dt}+
{\delta A\over\delta \theta_i^0}{d\theta_i^0\over dt}\right]=  \\
&=-\sum_{i}^N\left[\tau_2\sum_{k}^S \left({d\theta_i^k\over dt}\right)^2+\tau_3 \left({d\theta_i^0\over dt}\right)^2 \right]  \le 0. 
\end{split}
\end{eqnarray}Therefore whenever the system is falling or jumping into a new attractor the dynamics are effectively governed by the energy function, while during the protracted periods of permanence in the attractor, it is the adaptation function to be minimized, while the energy grows again, preparing for the next jump, as illustrated in Fig. \ref{fig:energia}. 

A set of simulations were conducted, however, considering what may be called a {\em rapidly adapting regime}, when in particular $\tau_3$ is very short, $\tau_3 \ll \tau_1 < \tau_2$, which has the computational advantage that rich dynamics unfold within limited CPU time. Approximate analytical considerations derived from those applicable to the slowly adapting regime can be extended to this rapidly adapting regime, which provides a sort of control case, although physiologically less plausible (it would model fast inhibition, but not its important slow component \cite{Koh+10}).

\subsection{Correlated patterns}\label{sec:corrpatt}

With the introduction of the adaptive thresholds, the network eventually slips away from any attractor it may have retrieved.
A free association is said to occur if the system then jumps into a new attractor. As shown in \cite{Rus+08}, such a jump is facilitated if there is a second memory pattern correlated with the first one. Indeed, while individual units are adapting and slowly changing their activation, those that show the same activation in both patterns act as a cue for the retrieval of the second memory.

To generate a set of correlated patterns we use the algorithm presented in \cite{Tre05}.
The algorithm is comprised of a two-step procedure. 
In the first step a number of mutually uncorrelated vectors, called \textit{factors}, are established. Each factor can influence the activation of some of the units, by "suggesting" a particular state.
In the second step the competition among these factors determines the final activation state of each unit in each of the $p$ patterns. 

In particular each factor acts on a distinct, but not exclusive, subset of $Na_f$ units with a strength that varies proportionally to the relevance of the factor itself and to a random number ranging from $0$ to $1$, which is nonzero with probability $a_{pf}$. The relative strength (relevance) of the factors decreases exponentially, on average, from the most relevant factor downward. The exponent $\zeta$ can be varied parametrically from a small value, producing nearly equivalent factors, to large ones, $\zeta\approx 1$, resulting in the dominant influence of the first few factors.
Each pattern has therefore a probability $a_{pf}$ to be influenced by a given factor. After all the relevant factors have expressed their suggestions on an activity direction among the $S$ possible ones, a full activation ($\sigma_i^k=1$) is assigned, in the state $k$ most activated by the factors, to the $Na$ most activated units (most activated in the selected direction), while the remaining $N(1-a)$ ones are set in the inactive state.
In this way, a set of $p$ patterns is produced with exactly $N(1-a)$ units in the inactive state and $Na$ units fully aligned with one of the $S$ active states.

\begin{figure}[ht]
\begin{center} 
\includegraphics[width=.45\textwidth]{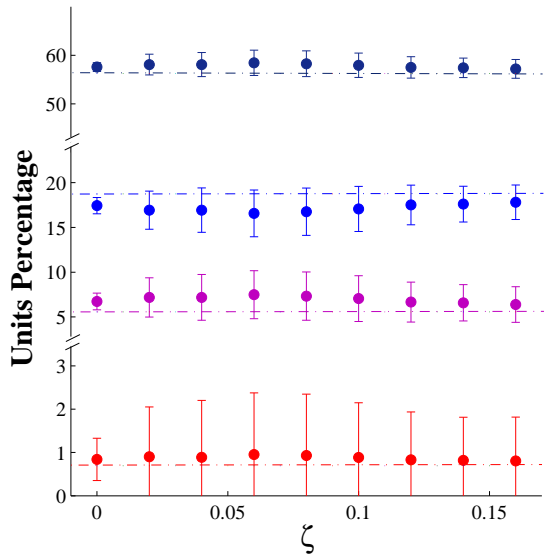}
\caption{(Color online) The generation algorithm produces a set of correlated patterns, whose correlation can be varied with the $\zeta$ exponent. With Potts units, where each can assume a variety of  active states, to gauge the correlation between different patterns (network configurations with a fixed number of active units) one has to take in account the number of units shared by the two patterns in the same active ($N_{as}$, red, bottom) and in the inactive ($N_{00}$, black, top) state; the number ($N_{ad}$, pink, third from top) of units active in two different states; and that ($N_{a0} = N_{0a}$, blue, second from top) of units active in a configuration and inactive in the other. The horizontal lines refer to the reference values of $N_{as}$, $N_{ad}$, $N_{a0}=N_{0a}$ and $N_{00}$, expressed as percentages, in the case of randomly generated patterns, with $a_{pf}=0.3$, $a=0.25$, $N=600$, $p=140$ and $S=9$. The effect of $\zeta$ on the average of these numbers is limited, as shown here in percentages, but intermediate values of $\zeta$ greatly amplify the variability of $N_{as}$ around its average.}
\label{fig:correl_apf_zeta}
\end{center} 
\end{figure}

The algorithm thus produces a set of patterns with a certain degree of correlation. The correlation can be tuned acting particularly on two parameters: $a_{pf}$ and the exponent $\zeta$ of the strength. Increasing the proportion of factors active in a pattern, $a_{pf}$,  
the average correlation between pairs of patterns increases almost linearly, as measured by the averages of the number of units active in the same state in both patterns, $N_{as}$, or in different states, $N_{ad}$, or both inactive in the pair, $N_{00}$, from the reference values taken for randomly generated patterns (not shown). 
The $\zeta$ exponent acts instead on the relative relevance of each factor. The downstream effect on the averages of $N_{as}$, $N_{ad}$ and $N_{00}$ is non-monotonic, as the average correlation increases with $\zeta$ for $\zeta$ small, but then it decreases; and its overall magnitude is limited anyway (as shown in Fig.\ref{fig:correl_apf_zeta}).  The effect is more salient on the variability of the smaller among these numbers, around their averages; in particular the average of $N_{as}$ is quite small (just above 4 units with the parameters of Fig. \ref{fig:correl_apf_zeta}), and its variability from pair to pair is considerably larger than the average. In other words, the algorithm produces patterns that {\em overall} roughly cover the same portion of their high-dimensional pattern space, with similar average correlations. But when a limited number of factors are relevant {\em local} relations are altered, with patterns typically having a neighborhood of other patterns with which they share many units active in the same state, and remote patterns with which they have no units in common, active in the same state. It is a non-transitive type of neighborhood (as the active units that define it vary from pair to pair), so it does not imply any clustering in pattern hyper-space.  

In \cite{Rus+08} we focus especially on the effect of the correlation between two patterns on the dynamics of a single latching transition. Here, instead, we aim to study how structural network parameters influence the dynamics of the whole latching sequence. Therefore, for the sake of simplicity, in most of the analyses we use a set of randomly generated patterns ($a_{pf}=0$ and $\zeta=0$), leaving to a brief section the discussion of how correlations alter the latching phase diagram. It is clear, however, that having more "neighbors" to any current pattern, with non-zero $a_{pf}$ and $\zeta$, facilitates latching transitions and also determines which transitions are likely to occur.

For random patterns, their average correlation is set by the sparsity $a$ and the number of active states $S$. Indeed, two patterns on average share $N(1-a)^2$ inactive units and they have $2Na(1-a)$ units active in just one of the two patterns, $Na^2/S$ units active in the same state and  $N(S-1)a^2/S$ units active in a different activation state (the reference values in Fig.~\ref{fig:correl_apf_zeta}).

\subsection{Storage capacity}\label{sec:storage}

A first observation, about the correlation between patterns, is that increasing it strongly affects the network retrieval performance, i.e. its storage capacity.

The capacity of the Potts network to store information, as with any neural network, is limited. For an autoassociative network, in particular, there is typically a maximal number $p_c$ of patterns, called its \textit{storage capacity}, that can be stored without significantly affecting the ability of the network itself to retrieve each of them from a partial cue \cite{Ami+87}. Exceeding this number, at the presentation of the cue the network, once driven solely by internal (recurrent) inputs, reaches a state uncorrelated with any of the memory patterns, or only partially correlated with the memory to be retrieved or, even, with a "wrong" one.

In \cite{Kro+05} the storage capacity $p_c$ is analyzed for a variety of Potts autoassociative networks.
For each of those, the capacity limit is studied as a function of the global parameter $a$, quantifying the sparsity of the memory patterns and indirectly of the network activity, and of the local parameter $S$, the number of Potts states. 

The network considered here is a graded-response and adapting variant of what in that analysis is called a "sparse" network (irrespective of activity becoming really sparse only in the limit $a \ll 1$). In the further "thermodynamic" limit $N, p, \beta \rightarrow \infty$, with $\alpha \equiv p/N$ finite, its storage capacity is found to scale, for $a \ll 1$, as


\begin{equation}
p_c \simeq \frac{N S^2}{4 a \ln(\frac{2 S}{a \sqrt{\ln(S/a)}})}
\end{equation}
for a fully connected network or, for a diluted network as

\begin{equation} \label{eq:storage}
p_c \simeq \frac{C S^2}{4 a \ln(\frac{2 S}{a \sqrt{\ln(S/a)}})},
\end{equation}
which are closer estimates than the simple $p_c\approx N S^2/a$ or $p_c\approx C S^2/a$, over a broad range of values of $a$, as they are for standard autoassociative networks \cite{Tre+91}.

This analytical result is validated by numerical calculations and by network simulations, but it only holds for the storage of random, uncorrelated patterns. One may now consider how an enhanced correlation among patters affects the storage capacity.

In a symmetric recurrent network in which one can define a Lyapunov or energy function, the correlation among memory patterns influences directly the energy depth of the network attractors, and hence its storage capacity. 
In a condition in which the stored patterns are few and uncorrelated, during the retrieval of a memory the overlap of the state of the system with that particular pattern is close to one, and close to zero with the others. 
In this case it has been shown \cite{Ami+85} that in the thermodynamic limit and for $T\to 0$ the attractor configurations of the network exactly match the stored patterns, and all have the same depth in energy. 
With correlated attractors, instead, when a memory is retrieved there is a partial activation of some of the other memories, roughly proportional to their correlation with the retrieved one. The energy minima then have a distribution in energy around the mean value.

Approaching the capacity limit one arrives to a condition where the basins of attraction of different memories overlap and eventually fracture and coalesce, forming new "spurious" stable states \cite{Rou+03}.
The main effect of correlations is similar, to generate overlaps among the attractors. This makes it more difficult for the network to retrieve the stored patterns.

\begin{figure}[ht]
\begin{center} 
\includegraphics[width=.45\textwidth]{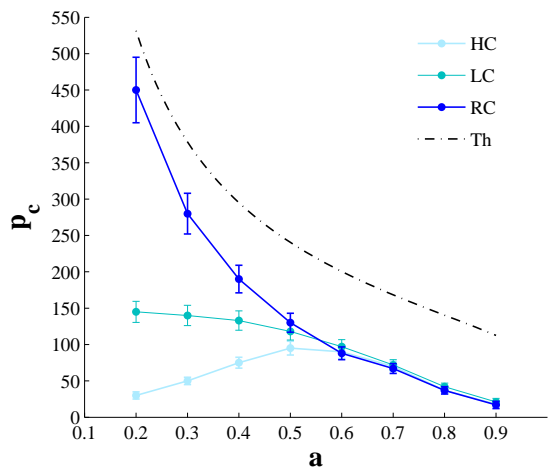}
\caption{(Color online) Storage capacity of a network of $N=600$ units with $S=5$ active states as a function of its sparsity $a$ for three sets of stored patterns: a set of randomly generated patterns ($RC$: $\zeta=0.0, a_{pf}=0.0$, top solid curve), one of weakly correlated ($LC$: $\zeta=10^{-6}, a_{pf}=0.4$, intermediate) and one of highly correlated ($HC$: $\zeta=0.1, a_{pf}=0.4$, bottom cuve). The dashed curve shows the theoretical value predicted by Eq.~\ref{eq:storage}, in the $a \ll 1$ limit. In these simulations, $U=0.5$, $w=0$, $C=90$ and $N=600$.}
\label{fig:storage}
\end{center} 
\end{figure}

Fig.~\ref{fig:storage} shows the difference in $p_c$ for sets of patterns with three different correlation levels, as obtained from computer simulations. From the figure, one sees that the presence of correlations, generated by the multi-factorial algorithm, decreases significantly the capacity of the net, when $a \to 0$, with respect to the randomly correlated case. One notes that, conversely, the effect of correlations is negligible for non-sparse patterns, $a\ge 0.5$, but in this range the actual capacity is sensibly lower than predicted by the asymptotic formula.

\section{Latching sequences}\label{sec:latch}

Once randomly or nonrandomly correlated patterns are stored, one may study the dynamics of the network elicited by cued retrieval. The initial retrieval is triggered here by a transient signal in the direction of a memory pattern, then the system is left to evolve subject to its own dynamics. What one may observe is shown in Fig.~\ref{fig:lungo}. After the first cued retrieval the system adapts and makes a transition to a new attractor. This new retrieved memory adapts in its turn, and acts as a cue for a new transition. 
Such a behavior is spontaneous and is due just to adaptation and to the correlation among the patterns. In these simulations, transitions are not due to noise since, at the level of interactions among Potts units, the dynamics is deterministic and no noise is included in the equations ($T$ represents instead noise internal to each patch, in the interactions among the neurons subsumed into the effective Potts description). At the same time the sequence of retrieved patterns is not defined {\em ad hoc}, and it does vary with any small variation of the initial conditions. Finally, one may notice that each of the retrieved patterns can appear multiple times in the retrieval sequence, as the increase in the thresholds is just temporary, and each unit can be reactivated and/or return to its former activity state, in due time.

Exploring the space of parameters, a variety of dynamical behaviors are qualitatively distinguished. For some parameters, that we call the \textit{finite latching region}, the complex dynamics, described above as latching transitions, die spontaneously after a certain number of \textit{steps} and the system relaxes into its inactive configuration, the only truly stable one for the network (Fig.~\ref{fig:lungo}(b)).
From this region, moving a bit in phase space, i.e. modifying the values of some of the parameters, it is possible to induce an increase or a reduction in the length of the latching sequence, and to approach either of two extreme conditions:
a \textit{no latching region}, where the system is able to recall the cued memory but not to express any associative transition after that (Fig.~\ref{fig:lungo}(a)),
and an \textit{infinite latching region}, where the latching process self-sustains indefinitely in time (Fig.~\ref{fig:lungo}(c)).

What makes latching begin, and what makes it eventually stop? 
We first present a qualitative description of the different regimes, or phases, and then analyze their boundaries, or phase transitions, in the simplified situation in which the stored patterns are uncorrelated or, to be exact, just randomly correlated. Structured correlations are then reintroduced by means of computer simulations.

\subsection{The onset of latching}

In \cite{Rus+08} we focus on the dynamics of a single latching transition, in order to clarify the internal mechanisms that enable the retrieval of a second pattern and to study qualitatively the variety of transitions arising from these dynamics. The analysis shows that, apart from the somewhat pathological case of oscillations between nearly overlapping attractors, latching transitions may be distinguished between random ones, dubbed \textit{low transitions} for the low value of the overlap between the latched patterns, and those driven by positive correlations, \textit{high transitions}.

\begin{figure}[ht]
\begin{center} 
\includegraphics[width=.5\textwidth]{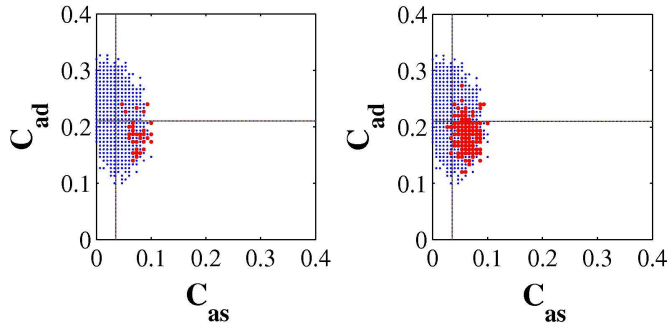}
\caption{(Color online) Latching transitions (darker, red dots) increase rapidly when the local feedback coefficient increases from $w=0.35$ (left) to $w=0.45$ (right). In either case, in the $C_{as}-C_{ad}$ correlation plane, transitions occupy a sub-region of that spanned by correlations among all the stored attractors (lighter, blue dots): they tend to occur only between pairs of patterns with a fraction of active units in the same state above the average value $C_{as}=a/S$ (vertical line) and a fraction in different states below the average value $C_{ad}=a(S-1)/S$ (horizontal line). In these simulations, $a=0.25$, $S=7$, and the stored patterns were only randomly correlated.}\label{fig:iniz}
\end{center} 
\end{figure}

Following this line, we produce $p$ retrieval sequences by initially cueing each time a different stored attractor. Fig.~\ref{fig:iniz} shows, for those cases in which
a transition is produced, the correlation parameters $C_{as}$ and $C_{ad}$ between pairs of latched patterns (red dots) with respect to all the pairs (blue dots). $C_{as}$ is the fraction of units, among the $Na$ active ones in the first pattern, that are also active in the second and in the same state, while $C_{ad}$ is the fraction of the same units that are also active, but in a different state; the complementary fraction $1-C_{as} -C_{as}$ are not active in the second pattern of the pair. One can see that, in agreement with \cite{Rus+08}, the transitions cluster in a region of the correlation plane. With these parameters (in particular, a relatively high threshold $U$), despite the presence of many patterns sharing a low value of $C_{as}$, no "low" transitions are observed. The percentage of patterns that give rise to a latching transition clearly increases going deeper into the finite latching region (Fig.~\ref{fig:iniz}(b)).

\begin{figure}[ht]
\begin{center} 
\includegraphics[width=.5\textwidth]{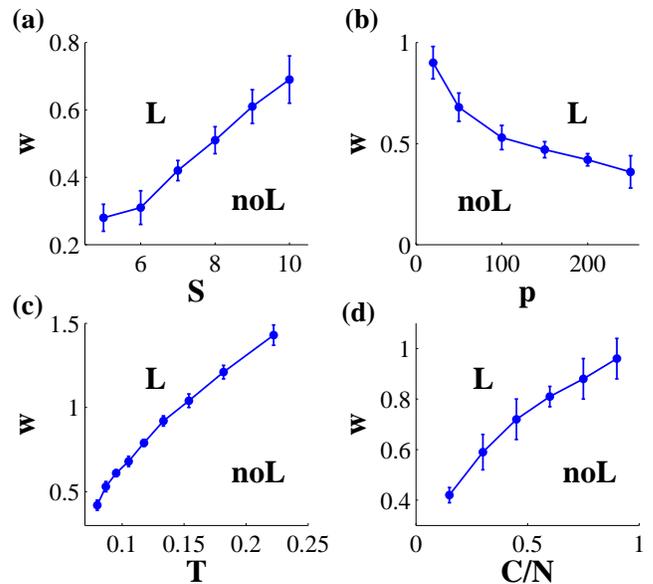}
\caption{The minimal value of the local feedback coefficient $w$ required to start latching is found with simulations to grow with $T$ (sub-linearly) and with $S$ (linearly or slightly supra-linearly); and to grow approximately with $\sqrt{C/p}$. The storage load ($p$) thus facilitates latching, whereas local noise, the number of states and the connectivity all obstruct it. When not varied explicitly, the parameters are set at $p=200$, $S=7$, $C=90$, $N=600$ and $T=0.08$.}\label{fig:inizio4}
\end{center} 
\end{figure}

Rather than asking whether transitions occur with a given correlation between the latched patterns, however, we want to focus here on how their occurrence depends on the parameters of the network. In order to study more systematically the border between the no latching and the latching regions, in particular, we vary the number of active states $S$, the number of stored patterns $p$, the network connectivity $C$ and the effective temperature $T$, and extract from simulations the lowest value of the local feedback term $w$ that enables latching transitions (Fig.~\ref{fig:inizio4}). Note that if these structural parameters are such that the network straddles the no latching and the finite latching region, transitions may occur or not, for the same parameters and even for the same set of patterns, depending e.g. on the initially cued configuration. This variability generates the error bars in Fig.~\ref{fig:inizio4}.

\subsection{Finite latching}

To better understand how network parameters affect sequence length, we fix a set of parameters and let the system evolve until it reaches a quiescent condition. If, after $6 \cdot 10^5$ updates of the whole network, equivalent to a time $1.8\cdot10^5~\tau_1$, the system is still active, the simulation is terminated.

\begin{figure}[ht]
\begin{center} 
\includegraphics[width=.5\textwidth]{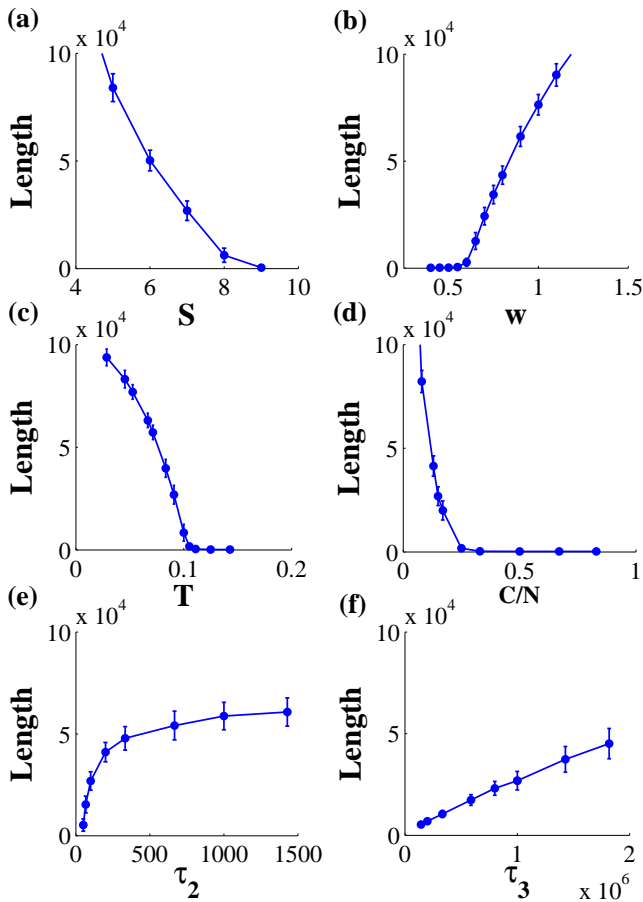}
\caption{Sequence length (in units of $\tau_1$) as a function of various parameters. The connectivity $C$, number of attractor states $S$ and local noise $T$ are confirmed to obstruct latching, whereas larger values of the local feedback $w$ (and storage load $p$, see Fig.~\ref{fig:distETVj}(a)) prolong it. Moreover, the sequence is found to be stretched almost in direct proportion to the general adaptation time scale $\tau_3$. When not varied explicitly, the parameters are set at $p=200$, $S=7$, $C=90$, $N=600$, $T=0.09$, $w=0.8$, $\tau_1=3.3$, $\tau_2=100$ and $\tau_3=10^6$.}
\label{fig:andamLungh}
\end{center} 
\end{figure}

Fig.~\ref{fig:andamLungh} shows the average length of the sequences so generated, in relation to the variation of several network parameters: $w$, $S$, $C$, $T$, $\tau_2$ and $\tau_3$, while the dependence on $p$ is shown in the next Figure. 

Focusing on the parameters defining network architecture, it turns out that an increase in memory load (Fig.~\ref{fig:distETVj}(a)), a reduction in the number of states of each unit (Fig.~\ref{fig:andamLungh}(a)) and a reduction in the connectivity (Fig.~\ref{fig:andamLungh}(d)) all give rise to an increase in sequence length.
Such increase is also produced reducing the effective temperature $T$ (Fig.~\ref{fig:andamLungh}(c)) as well as increasing the local feedback term $w$ (Fig.~\ref{fig:andamLungh}(b)). It is important to note that, for each of these parameters, there is a critical value beyond which no latching is observed -- and often another value, as seen in Fig.~\ref{fig:distETVj}, beyond which latching appears to proceed indefinitely.

Finally, simulations indicate (Fig.~\ref{fig:andamLungh}(e, f)) that, while the sequence length tends to a plateau by increasing the specific adaptation time scale $\tau_2$, there is a monotonic, quasi-linear dependence with the generic threshold adaptation time constant, $\tau_3$.

\subsection{Within the finite latching region}\label{sec:phas-ton}

Once the parameter set is fixed, the length of the latching sequence is still not entirely defined. There is indeed variability due to the initial conditions. As previously mentioned, it is enough to slightly jitter the cued configuration to change completely the sequence of retrieved patterns. While considering the dependence on the memory load $p$, we focus then on the distribution of lengths seen even for fixed parameters, and due to different initial conditions.

In order to explore the whole configuration landscape, after the storage of the memory patterns, we cue the network in the direction of each of the $p$ patterns, in turn.

\begin{figure*}[ht]
\begin{center} 
\subfigure{\includegraphics[width=.48\textwidth]{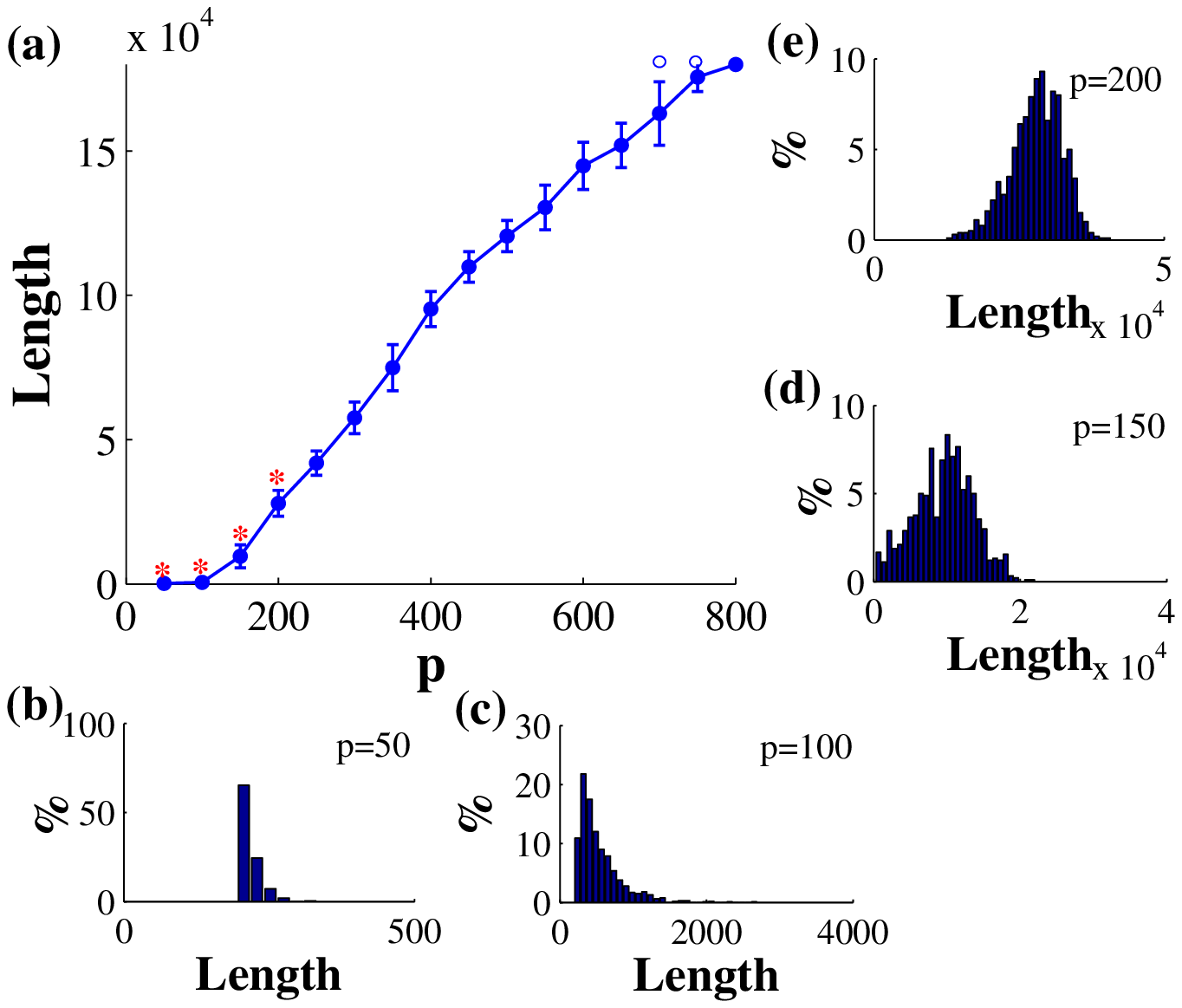}}
\hspace{5mm}
\subfigure{\includegraphics[width=.48\textwidth]{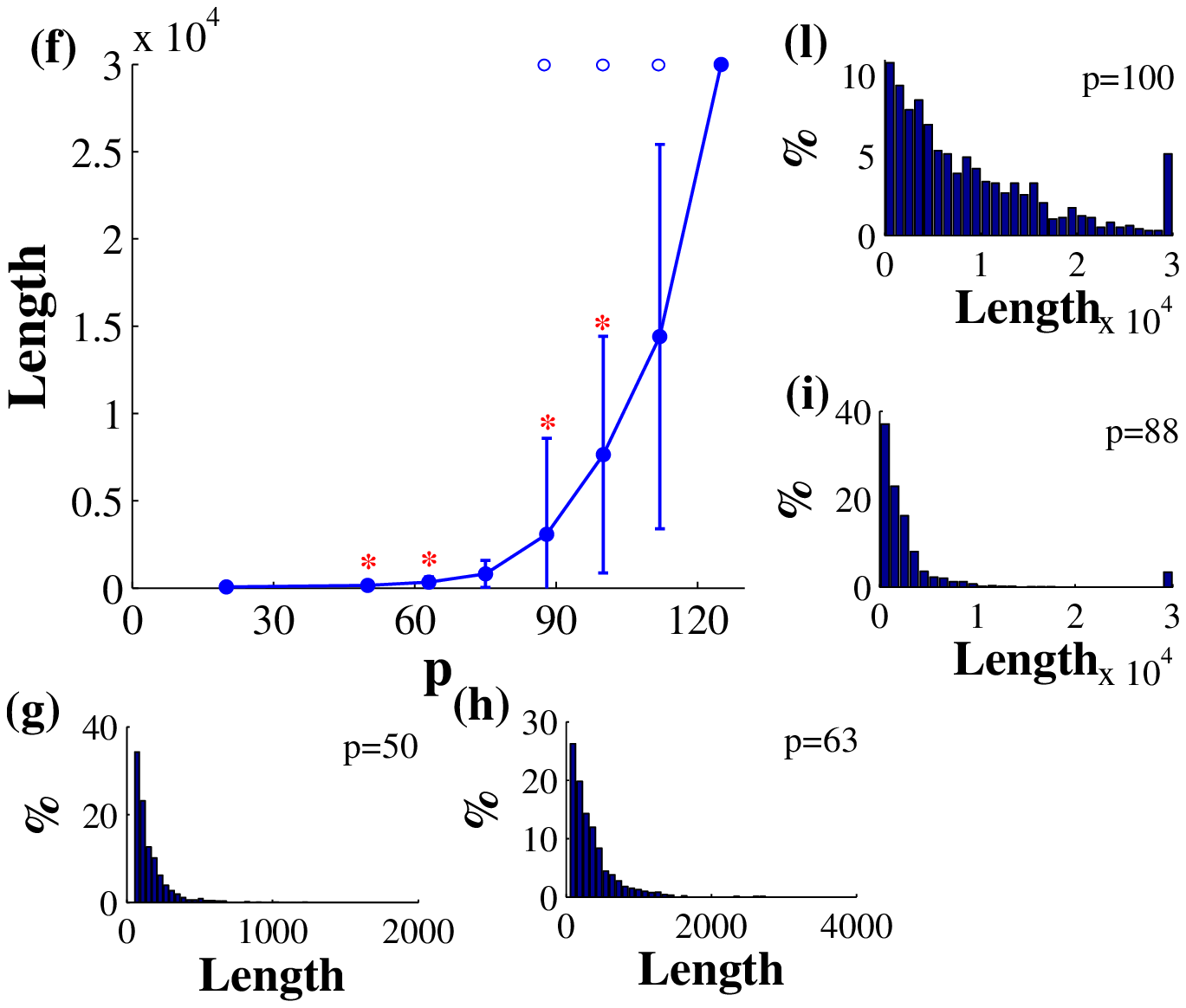}}
\caption{Sequence length, in units of $\tau_1$, with respect to the number $p$ of stored patterns when the network is in a slow adaptation regime ($\tau_1= 3.3$, $\tau_2= 100$, $\tau_3= 10^6$) (a) and in a fast adaptation regime ($\tau_1= 20$, $\tau_2= 200$, $\tau_3= 10$) (f). (b, c, d, e, g, h, i, l) Distribution of sequence lengths for a fixed $p$ (at the values indicated by the stars). In the slow adaptation regime, at latching onset, the distribution of lengths is strongly skewed towards low values, with a tail approaching an exponential or Poisson shape (b, c); as $p$ grows and the mean length increases, the distribution is increasingly normal (d, e). The crossover appears to occur when the mean length starts to increase strongly, and quasi-linearly, with $p$. In the fast adaptation regime, instead, one portion of the distribution remains quasi-exponential (g, h, i, l), but the error bars increase for $p>80$, due to the inclusion in the distribution of an increasing fraction of trials with an undetermined length, because they were terminated upon reaching the maximum allowed duration of $3\times 10^4\tau_1$ (the occurrence of such trials is denoted by empty circles; allowing for longer duration did not increase the quasi-exponential portion appreciably, given $p$). }
\label{fig:distETVj}
\end{center} 
\end{figure*}

In Fig.~\ref{fig:distETVj} one can see the distribution of sequence lengths as the network moves from the no latching condition towards the indefinite latching region.
Emerging from the no latching phase, as $p$ increases the network begins to latch, initially with only a few jumps between attractor states (the mean sequence length is so short that it is hard to distinguish it from the $x$-axis). In this condition, the variability in sequence length is spread with a mode close to zero (the observed gap of 200 time steps is approximately the time for the first retrieval to be externally triggered) and a roughly exponential tail (Fig.~\ref{fig:distETVj}(b, c)). As the average length increases, however, the distribution moves from the exponential to a more normal shape (Fig.~\ref{fig:distETVj}(d, e)), suggesting a simple relaxation process with a characteristic time scale, which turns out to be just $\tau_3$ (Fig.~\ref{fig:andamLungh}(f)). During this relaxation the system retains a simple memory of the time spent latching, residing in the mean value of the generic adaptation threshold, $\theta^0$. With time constant $\tau_3$, in fact, $\theta^0$ approaches its asymptotic value, eventually terminating the latching sequence, when on its way it crosses a certain critical value.

Although so far we have discussed the slowly adaptive regime of Sec.~\ref{sec:slowregime}, the complexity seen in Fig.~\ref{fig:distETVj}(a, b, c, d, e) makes it now convenient to review the sequence length distribution in the {\em rapidly adaptive regime}. As explained in Sec.~\ref{sec:Adaptivedynamics}, in such a regime $\tau_3$ is a fast time constant which can be interpreted as characterizing the immediate regulation of the mean activity level locally in the cortex, by fast GABA$_A$ inhibitory mechanism. In such a regime, as prescribed by Eq.~\ref{eq:eqb3}, the threshold $\theta_i^0$ imposed on each Potts unit tracks the activation level $\sum_{k=1}^S\sigma_i^k$ of the unit itself with speed $1/\tau_3$. Then, by the time the next latching transition occurs, the threshold almost equals the unit activation itself, and it is removed as a distinct variable, thus simplifying the dynamics of the system.

Despite the increase in the sequence lengths, common to both regimes, in the rapid adaptation regime we find a simpler scenario. For a fixed $p$ the length distribution appears quasi-exponential (Fig.~\ref{fig:distETVj}(g, h, i, l)). This trend is characteristic of an essentially "memory-less" process, in which at each step the probability to fall into the quiescent state, and terminate the sequence, is roughly the same. For higher $p$, (Fig.~\ref{fig:distETVj}(i, l)), one notes the appearance of potentially "infinite" sequences, although we terminate them after $3 \times 10^4$ steps, thus making the average length finite. For the same sets of patterns, depending on the initial condition, hence on the landscape visited by the dynamics, finite and infinite (manually terminated) latching sequences co-occur. The increase in the average length for higher $p$ is then due both to an increase in the length of finite sequences and to an increase in the proportion of terminated sequences (not shown).

The quasi-exponential length distribution earlier observed in the slow adaptation regime close to the no-latching region might have been interpreted as resulting from the tail of a normal distribution, when its mode would be below zero, after its negative component is cut off or squashed towards zero.
On the other hand, in the fast adaptation regime, the quasi-exponential tail is all that is seen, because when latching does not stop in a few steps, it proceeds indefinitely. But what can terminate it in a few steps, in either regime? The large variability indicates a process, in which the network sometimes falls in a "trap", i.e. in the basin of attraction of one or few "isolated" memory patterns, out of which it cannot latch further. 

To test this hypothesis, for a fixed set of patterns we have run extensive simulations, keeping track of the last attractor visited before latching subsides and the network falls in the globally quiescent state. 

\begin{figure}[ht]
\begin{center}
\includegraphics[width=.5\textwidth]{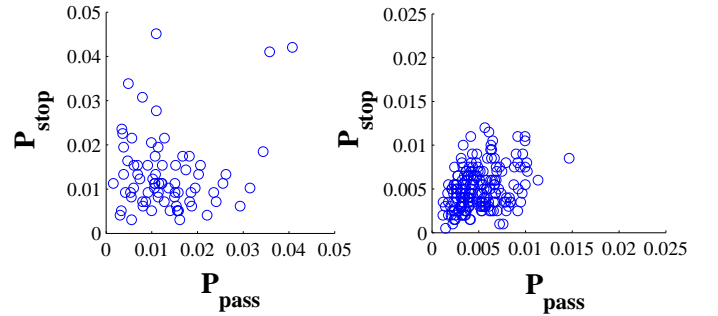}
\caption{Probability of a pattern to be the last of a latching sequence ($P_{stop}$) vs. its probability to occur at any position ($P_{pass}$) for the fast (left, $p=100$) and slow (right, $p=200$) adaptation regimes.}
\label{fig:tuttip}
\end{center}
\end{figure}

Fig.~\ref{fig:tuttip} shows for each pattern the probability to be the last in a latching sequence ($P_{stop}$), with respect to its probability to occur at any position in a sequence ($P_{pass}$).
In the slow adaptation regime of Fig.~\ref{fig:tuttip}, right the patterns have a similar probability to occur as the last of a sequence. In contrast, in the fast adaptation regime of Fig.~\ref{fig:tuttip}, left there seem to be some "preferred" patterns where latching often terminates.
Since those patterns are not apparently more likely to be visited than others before a sequence ends (the relative $P_{pass}$ is not higher than for other patterns) there must be something in their correlational surround that makes them less able to cue the retrieval of other attractors, so that dynamics often stop there. These considerations suggest that the transition to infinite latching occurs when from each basin of attraction a "border crossing" is open to another basin of attraction that the system can latch to, when slipping away of the former. 
\subsection{Sustained latching}\label{sec:sustained}

\begin{figure}[ht]
\begin{center} 
\includegraphics[width=.5\textwidth]{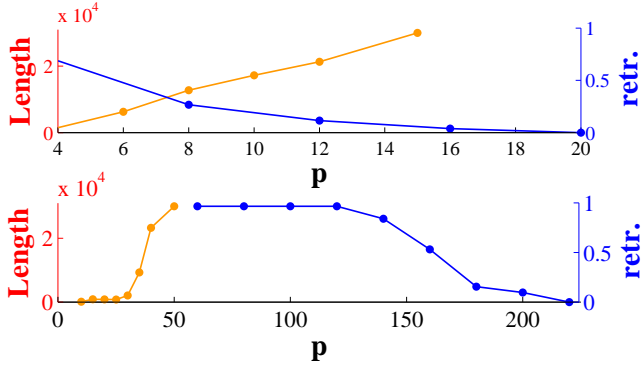}
\caption{(Color online) Average sequence length (orange ascending curve; left scale, in units of $\tau_1$) and retrieval quality (blue descending curve; right scale) as a function of the storage load, for $S=3$ (above) and $S=7$ (below).}
\label{fig:finestra}
\end{center}
\end{figure}

As introduced in the previous paragraph, moving further in parameter space, beyond the finite latching region, one seems to enter a domain of infinite sequences, where each latching step is followed by another one, until the simulation has to be terminated. 
It is important to note that while approaching such infinite latching, the network is also moving towards its storage capacity limit. For some values of the parameters, indeed, the capacity limit, although somewhat ill-defined in a network of moderate size, appears to be crossed before the infinite latching region is reached. If this happens, the system sustains an indefinite noisy activity, without ever stopping and without fully retrieving any pattern. For example, Fig.~\ref{fig:finestra} shows the average sequence length, if not terminated at $3\times 10^{4}~\tau_1$, and the faction of patterns retrieved with an accuracy higher than $90\%$ (a measure of retrieval quality) if network dynamics are initiated by a partial cue.
When $S=7$ one can see that the system reaches the infinite latching region around $p=50$, while retrievability decreases beyond $p=120$. Therefore there is a window in the storage load range, in which the system really latches from a retrieved memory to the next, indefinitely.
When $S=3$, instead, the ability to retrieve a cued pattern starts to decline already around $p=8$, when the length of the latching process is still finite and in fact short. Therefore, for $S=3$ and with this set of parameters, the network does not ever really latch indefinitely, but rather enters a region of indefinitely protracted noisy dynamics.

\subsection{Rapid adaptation: a percolation type of transition}\label{sec:percolation}

The observations of Sec.~\ref{sec:phas-ton} suggest that latching dynamics stop when the network runs into an attractor with no "border crossings", or links, to other basins of attraction. Exploring this hypothesis is, however, complicated by the presence of adaptation that continuously modifies the basins and their boundaries. 

We consider here the rapidly adapting regime, $\tau_3 \ll \tau_1 < \tau_2$, in which the generic threshold adapts so rapidly that it can be considered as a function quasi-instantaneously set to the value imposed by the $\{\sigma\}$'s, as if due to rapid inhibition. In this case, the effective energy minimized during the rapid transition dynamics does not depend on the $\theta^0$'s any longer, only the $\theta^k$'s for $k\ne 0$ enter as parameters, and the dependence on the $\sigma^k$'s is modified by an initial term 
\begin{widetext}
\begin{eqnarray}
\begin{split}
E =& \frac{1}{2} \sum_i^N(\sum_{k=1}^S \sigma_i^k)^2 -{1\over 2}\sum_{i,j\ne i}^N\sum_{k,l=1}^S J_{ij}^{kl}\sigma_i^k\sigma_j^l - {w\over 2}\sum_i^N\left[\sum_{k=1}^S(\sigma_i^k)^2 - {1\over S} (\sum_{k=1}^S\sigma_i^k)^2\right]+ \\
& +\sum_i^N\sum_{k=1}^S \left\{ (U+\theta_i^k)~\sigma_i^k + {1\over \beta} \left( \sigma_i^k \ln {\sigma_i^k\over \sigma_i^k+\sigma_i^0}+ \sigma_i^0 \ln {\sigma_i^0\over \sigma_i^k+\sigma_i^0}\right) \right\} \label{eq:Efast} 
\end{split}
\end{eqnarray}
\end{widetext}necessary to take into account the fast $\theta^0$ dynamics.

To get an impression of the energy landscape visited by the system, without the distortion of adaptation, we consider also $E_{land}$, that is, Eq.~\ref{eq:Efast} without the contribution of $\theta_i^k$ (while $\theta_i^0$ is kept, though hidden as a function of the $\{\sigma\}$'s).

\begin{figure}[ht]
\begin{center} 
\includegraphics[width=.5\textwidth]{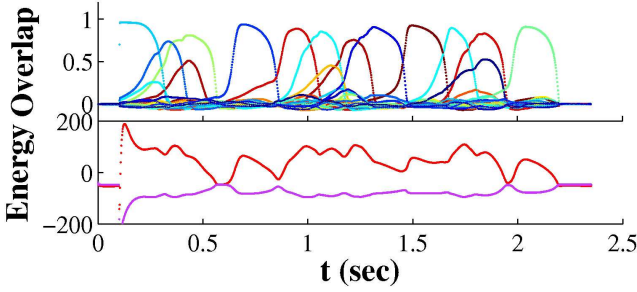}
\caption{(Color online) Latching transitions and relative value of the total energy ($E$, higher, red curve of the bottom graph) and of the landscape energy ($E_{land}$, lower, purple curve). At each attractor transition, the energy decreases while $E_{land}$ has a corresponding peak.}
\label{fig:dueEnergie}
\end{center} 
\end{figure}

Fig.~\ref{fig:dueEnergie} shows the overlaps of the system with the stored patterns and the two energy values as a function of time (expressed in {\em sec}, assuming $\tau_1=10 msec$).
In agreement with Sec.~\ref{sec:Adaptivedynamics} there is a drop in the energy (red line) every time the system has a latching transition, however noisy and far from the ideal of allowing only two non-zero overlaps the transition may be. 
At the same time, though, $E_{land}$, the energy related to the landscape (purple line), has a peak. This suggests the presence of an energy barrier that has to be passed each time the system changes basin of attraction. If this hypothesis is correct, we expect the increase in sequence length, approaching the infinite latching region, to be due to a decrease in such barriers.
We run then new simulations, crossing the transition to infinite latching by increasing the number of stored patterns.

\begin{figure}[ht]
\includegraphics[width=.5\textwidth]{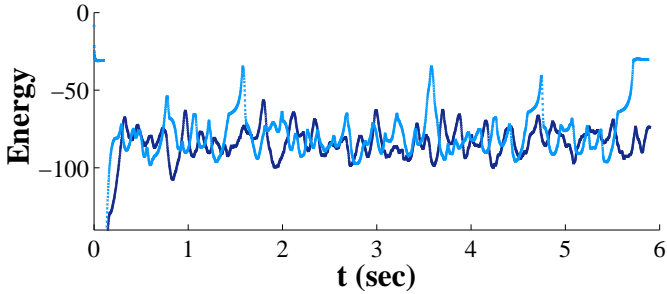}
\caption{(Color online) Landscape energy profiles ($E_{land}$) for a network with load $p=20$ (light blue) and $p=80$ (darker blue). The increase in the number of stored patterns reduces the highest energy boundaries between attractors.}
\label{fig:2080}
\end{figure}

Fig.~\ref{fig:2080} shows the energy profile of the system when set below its infinite latching transition (light blue) and above (blue). Comparing the two curves, no major difference is visible between the energy levels of the attractors, i.e. the bottom of the valleys. This appears to exclude a scenario in which the ability to produce an infinite sequence is due to the presence of deeper minima. 
On the other side, by increasing the number of stored patterns, the energy boundaries which separate two attractors decrease, especially the highest ones. This confirms our hypothesis and suggests that what actually makes the system stop in a particular attractor is the height of the (landscape) energy boundaries surrounding it.

We can then imagine the energy landscape of the system as a high-dimensional manifold, where the presence of stored patterns is marked by valleys, that correspond to their basins of attraction.

\begin{figure}[ht]
\begin{center} 
\includegraphics[width=.5\textwidth]{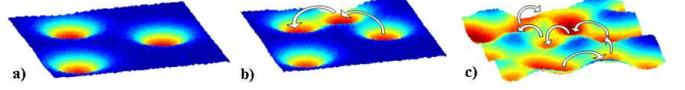}
\caption{(Color online) 2-dimensional sketches of the energy landscape. (a) When just a few attractors are stored, the system retrieves memories but it is not able to latch. (b) If the number of attractors increases, the basins connect one to the other, allowing some latching transitions. (c) A further increase in the number of stored patterns leads to a condition in which, irrespective of the starting point, the dynamics go on indefinitely in time.}
\label{fig:landscapes}
\end{center} 
\end{figure}

When a few sparse attractors are stored (Fig.~\ref{fig:landscapes}(a)) the network, if cued towards one of those, works correctly as a memory but without performing any latching transition. The large distance among patterns, indeed, does not allow latching to a new attractor, and once exited from the first attractor the network remains permanently in its stable inactive state (Fig.~\ref{fig:lungo}(a)).

If we add new patterns, however, some of their basins of attraction connect (Fig.~\ref{fig:landscapes}(b)). This lowers the energy boundaries enough, allowing the system to make transitions and creating privileged paths in the landscape. 
A similar condition is visible in Fig.~\ref{fig:lungo}(b), where the system shows latching dynamics, that remain however limited in time. 
The refractory period due to adaptation, indeed, impedes a new immediate recall of the first memory after the retrieval of the second one.
In order to have consecutive transitions, an attractor needs then at least two other correlated attractors, the previous and the following one. 
The probability of such situation to occur increases with the number of stored attractors, and its inherent variability produce a consequent variation in the length of the latching sequence.

By adding new valleys we arrive finally to a state in which, starting from whatever point in the manifold, the system can always find an infinite path across the landscape (Fig.~\ref{fig:landscapes}(c) and Fig.~\ref{fig:lungo}(c)). This behavior is typical of a {\em percolation phase transitions} \cite{Sta+85,Pha+84}. The increase in the probability 
of a border crossing, which is higher the more crowded and the wider are the basins corresponding to different memories, and which increases the average sequence length, can then be regarded as the probability of a link in a graph, where the nodes are the attractors. Infinite latching might then require, as in a percolation phase transition, this probability to be above a certain critical value (leading to the formation of a so called {\em giant component}).

\section{An approximate analytical approach}\label{sec:analytical}

In order to understand the emergence of distinct phases, let us consider the time derivative 
of the energy, Eq.~\ref{eq:energy}, when the network is still close to a given attractor $\vec{\xi^{\mu}}$ and a single unit is changing state, pretending to be in an ideal situation with only clean latching transitions from a single active attractor to the next. We stay now within the slowly adaptive regime. We focus first on a unit $i$ which is currently inactive, i.e. $\xi^{\mu}_i=0$
and $\sigma_i^0\simeq 1$, at the moment just before it starts to increase its activation in a particular direction $k$, so that a moment later $d\sigma^k_i / dt >0$. Clearly, the change will lead to a decrease in energy if $dE_i/d\sigma^k_i<0$, and in that case it is favoured, provided the thresholds can be assumed to stay roughly constant. When still at equilibrium, however, $dE_i/d\sigma^k_i=0$ (in all $S$ directions), and in fact the small activation values $\sigma_i^k\simeq 0$ of the unit are determined by imposing that all such first derivatives vanish.

Considering that $\sigma^0_i=1 -\sum_l \sigma^l_i$, the first derivatives are calculated as
\begin{eqnarray}
\begin{split}
{dE(\{\sigma^m_i\})\over dt} &= \sum_{k=1}^S{\delta E(\{\sigma^m_i\})\over \delta \sigma^k_i}{d \sigma^k_i\over dt} =\\
=& \sum_{k=1}^S\left[{\delta E_i(\{\sigma^m_i\},\sigma^0_i)\over \delta \sigma^k_i}-{\delta E_i(\{\sigma^m_i\},\sigma^0_i)\over \delta 
\sigma^0_i}\right]{d \sigma^k_i\over dt} 
\end{split}
\end{eqnarray}
\begin{eqnarray}
\begin{split}
&{\delta E(\{\sigma^m_i\})\over \delta \sigma^k_i}=-\sum_{j\ne i}^N\sum_{l=1}^SJ_{ij}^{kl}\sigma^l_j-w(\sigma^k_i-{1\over S}\sum_{l=1}^S\sigma^l_i)+\\
& +U+\theta^0_i+\theta^k_i+T\left(\ln {\sigma^k_i \over \sigma^k_i+\sigma^0_i} -\sum_{l=1}^S \ln {\sigma^0_i \over \sigma^l_i+\sigma^0_i}\right). \label{eq:dEidt}
\end{split}
\end{eqnarray}
The terms after the first, in the partial derivatives, include some (the thresholds) that do not depend on $\sigma^k_i$, some (those 
proportional to $w$) that are only linear in $\sigma^k_i$, and the {\em entropy} terms, 
proportional to $T$, that are highly nonlinear in $\sigma^k_i$, going from $-\infty$ to $+\infty$ as $\sigma^k_i$ goes from 0 to 1. The first term, that reflects the coupling to the other units, $-\sum_{j\ne i}^N\sum_{l=1}^SJ_{ij}^{kl}\sigma^l_j$, is constant with respect to the activation values and thresholds of unit $i$, but it varies a great deal from unit to unit, even among those we are focusing on, inactive in the current configuration and which are activating if a transition occurs. It can be estimated through a signal-to-noise analysis, by considering separately in the couplings $J_{ij}^{kl}$ the contributions due to the storage of the current memory attractor, $\mu$, of the attractor after the transition, $\nu$, and of all others, $\varrho\ne\mu,\nu$. The signal (largely consistent across those units) is due to the first two, and its average is denoted as $\bar{J\sigma}$, while the variability is due mainly to the storage of the other memories, and to the incomplete connectivity, $c_{ij}\ne 1$, which makes the signal (and noise) vary depending on which are the exact presynaptic units. The standard deviation of this variability is denoted here as $\Delta_{J\sigma}$.

\subsection{The signal and noise from the network}

The signal may be estimated as
\begin{eqnarray}
\begin{split}
-\langle\sum_{j\ne i}^N\!\sum_{l=1}^SJ_{ij}^{kl}\sigma^l_j\rangle \equiv& -\bar{J\sigma}\simeq {a \over S}\! -\!\left[\Gamma_{as}\!-\!\Gamma_{ad}\right]{a^2\over S^2}(S-1)+ \\
&-\left[\Gamma_{as}-\Gamma_{a0}\right]{a\over S}(1-a) \label{eq:signdE}
\end{split}
\end{eqnarray}
where $\Gamma_{as}, \Gamma_{ad}$ and $\Gamma_{a0}$ measure the actual correlation between the current and the forthcoming attractor, relative to their expectation values if they were randomly correlated. Specifically, they denote the ratios between the number of units presynaptic to the given unit that are active in the current attractor and are active in the same state ($\Gamma_{as}$), in a different state ($\Gamma_{ad}$) or inactive ($\Gamma_{a0}$) in the forthcoming attractor, and their network averages, which can be estimated as follows (in the case of randomly correlated patterns)
\begin{eqnarray}
 \Gamma_{as} &=& N_{as}/< N_{as}> \nonumber\\
 \Gamma_{ad} &=& N_{ad}/< N_{ad}>\nonumber\\
 \Gamma_{a0} &=& N_{a0}/< N_{a0}> \nonumber
\end{eqnarray}
\begin{eqnarray}
<N_{as}>&=&Na<C_{as}>=Na^2/S\\
<N_{ad}>&=&Na<C_{ad}>=Na^2(S-1)/S\nonumber\\
<N_{a0}>&=&Na <1-C_{as}-C_{ad}>=Na(1-a).\nonumber \label{eq:gammas}
\end{eqnarray}
Therefore on average $\Gamma_{as}=\Gamma_{ad}=\Gamma_{a0}=1$ and the only signal comes from the current attractor (the first term in Eq.~\ref{eq:signdE}). As illustrated in Fig.~\ref{fig:iniz}, however, latching transitions are favoured among attractors with $\Gamma_{as} > 1$ and $\Gamma_{ad}, \Gamma_{a0} < 1$, since these values bring down the energy derivative. 
In particular the major contribution is due to the $\Gamma_{as}$ term (which is the one that most deviates from unity when correlated patterns are generated by the multifactorial algorithm, see Fig.~\ref{fig:correl_apf_zeta}). For $p$ random patterns, the one with the largest number of shared active units in the same state as the current attractor can be estimated, from the tail of a Poisson distribution, to have an excess number that roughly satisfies the equation
\begin{equation}
\ln p = < N_{as}> \left[ {N_{as}\over < N_{as}>} \ln {N_{as} \over < N_{as}>e} + 1\right].
\end{equation}
For example, if $\Gamma_{as}\simeq 3$ and $\Gamma_{ad} \simeq \Gamma_{a0}\simeq 1$, we have $-\bar{J\sigma}\simeq -a / S$,
that is, the excess correlation with the forthcoming attractor is sufficient to switch the sign of the signal from positive (resisting change) to negative (favouring change).
 
The variability across units is due to several sources, the most important of which are the presence of the other stored patterns, and the incomplete connectivity. Both these factors generate terms that are on average zero across units, but with a variance that can be estimated to be, respectively,
\begin{eqnarray}
\begin{split}
&\langle \sum_j^N \!\frac{c_{ij}}{(C a(1-\frac{a}{S}))^2}\!\sum_{\varrho=1}^{p-2}\sum_{l=1}^S (\delta_{\xi_{i}^{\varrho}k}\! -\!\frac{a}{S})^2 (\delta_{\xi_{j}^{\varrho}l} \!-\!\frac{a}{S})^2 \delta_{\xi_{j}^{\mu}l} \rangle \simeq \\
&\simeq {a(p-2)\over CS^2}\\
\end{split}
\end{eqnarray}
\begin{eqnarray}
\begin{split}
&\left[\langle \sum_{j}^N\sum_{l=1}^S (c_{ij}\delta_{\xi_{j}^{\mu}l})^2 \rangle_{c_{ij}}- \sum_{j}^N \sum_{l=1}^S 
\langle c_{ij}\delta_{\xi_{j}^{\mu}l}  \rangle_{c_{ij}}^2\right] \left[{\bar{J\sigma} \over Ca}\right]^2 \simeq \\
&\simeq  {(1-Ca/N)\over Ca} \left(\bar{J\sigma}\right)^2 , 
\end{split}
\end{eqnarray}
which leads to a contribution, taking one standard deviation in the downward direction favouring the transition
\begin{eqnarray}
\begin{split}
&-\!\left\{\!\langle \left(\sum_{j\ne i}^N \sum_{l=1}^S J_{ij}^{kl} \sigma^l_j \right)^2 \!\rangle \!-\! \left(\langle\sum_{j\ne i}^N\sum_{l=1}^S J_{ij}^{kl} \sigma^l_j\rangle \right)^2 \right\}^{1/2} \!\!\equiv  \\
&\equiv -\Delta_{J\sigma} \simeq  \\
&\simeq  -\sqrt{{a(p-2)\over C S^2}+{a(1-Ca/N) \over C S^2}\left(S\bar{J\sigma}/a\right)^2} \simeq  \\
&\simeq -{\sqrt{a}\over S\sqrt{C}} \sqrt{(p-2)+\left(S\bar{J\sigma}/a\right)^2}. \label{noise_amp}
\end{split}
\end{eqnarray}

\subsection{The single unit terms}\label{sing_unit_term}
 
The variability in the coupling term across the $S$ directions also gives the different activation
values
\begin{equation}
\sigma^k_i\simeq\exp \beta\left(\bar{J\sigma}\pm \Delta_J -U-\theta^0_i-\theta^k_i\right)
\end{equation}
which are all small if the unit is largely inactive, with the signal $\bar{J\sigma}$ insufficient to overcome the threshold $U$. Given the exponential dependence on the fluctuations in the coupling, however, one activation will typically be much larger than the others, which can be taken to hover around their mean value $\bar{\sigma}\simeq 0$.
The stability to fluctuations that activate the unit further in the direction in which it is leaning is determined by the stability matrix
\begin{eqnarray}
\begin{split}
{\delta^2 E (\{\sigma^m_i\})\over \delta \sigma^k_i\delta \sigma^l_i}&\!=\!-w(\delta_{kl}\!-\!{1\over S})\!+\!T\!\left[\delta_{kl} \left({1\over \sigma^k_i}-{1 \over \sigma^k_i+\sigma^0_i}\right)+  \right. \\
&\left.   +\!{1 \over \sigma^k_i+\sigma^0_i}\!\!+\!{1 \over \sigma^l_i+\sigma^0_i}\!-\!\!\sum_{m=1}^S {1 \over \sigma^m_i+\sigma^0_i} +{S\over \sigma^0_i}\right].\label{eq:d2Eidt}
\end{split}
\end{eqnarray}
When the active states are all activated with the same amplitude, the stability matrix has one eigenmode corresponding to the increase of the common amplitude, which is safely stable, and $S-1$ degenerate eigenmodes corresponding to increasing amplitudes in each direction. 
When instead one state is already more active, and the others are close to $\bar{\sigma}$, one of the $S-1$ eigenmodes becomes unstable earlier, and it is the one proportional to $\delta_{1k}+(\delta_{1k}-1)\epsilon^k/S$, where we assume an instability in the direction 1
and the small corrections $\epsilon^k \propto \sigma^k$ for $k\ne 1$ express the difference between considering the full stability matrix and simply the second derivative
$\delta^2 E_i(\{\sigma^m_i\})/ (\delta \sigma^1_i)^2$.
The eigenvalue $\lambda$ satisfies the equations (dropping the unit index $i$, and for all $k\ne 1$)
\begin{eqnarray}
\begin{split}
\beta\lambda=&-\beta w (1-{1\over S})+{1\over \sigma^1}+{S\over \sigma^0}
-\sum_{l\ne 1}^S{1\over \sigma^0+\sigma^l}+  \\
&-\sum_{l\ne 1}^S{\epsilon^l\over S}\left[\beta{w\over S}+{S\over\sigma^0}- 
\sum_{m\ne 1,l}^S{1\over \sigma^0+\sigma^m}\right]\label{eq:eigen}
\end{split}
\end{eqnarray}
\begin{eqnarray}
\begin{split}
-\beta\lambda{\epsilon^k\over S}=&\,\beta {w\over S} + {S\over \sigma^0}
-\sum_{l\ne 1,k}^S{1\over \sigma^0+\sigma^l}+{\epsilon^k\over S}(\beta w-{1\over \sigma^k}) +  \\
&-\sum_{l\ne 1}^S{\epsilon^l\over S}\left[\beta{w\over S}+{S\over\sigma^0}- 
\sum_{m\ne l,k}^S{1\over \sigma^0+\sigma^m}\right]
\end{split}
\end{eqnarray}where the equation after the first can be taken to determine the $\{\epsilon^k\}$'s as a Taylor expansion in the $\{\sigma^k\}$'s, while the first determines the eigenvalue $\lambda$ to zero'th order in the expansion, as a function of $\sigma^1$, and, combined with the other equations, the following orders. 

At the instability, $\lambda = 0$. This determines $\epsilon^k $ to be, to first order in $\sigma^k$,
\begin{equation}
\epsilon^k \simeq \sigma^k \left( \beta w +2S/\sigma^0 \right) + O(\bar{\sigma}^2)
\end{equation}
and gives the further condition
\begin{eqnarray}
\begin{split}
\beta w \left(1 - {1\over S}\right) \simeq & {1\over \sigma^0} + {1\over\sigma^1}+\\
-{1\over (\sigma^0)^2}&\sum_{l\ne 1}^S\sigma^l\left[ \left(2+{\beta w \sigma^0 \over S}\right)^2-1\right] + O(\bar{\sigma}^2)  \label{eq:dE2}
\end{split}
\end{eqnarray}
that has to be satisfied together with $\delta E_i/\delta \sigma^1=0$ and, in principle, $\delta E_i/\delta \sigma^k=0$ for $k\ne 1$ to yield $\sigma^1, \sigma^k$ and the dependence of e.g. $w$ on all other parameters at the phase transition. In practice, if one can approximate $\sigma^k\simeq 0$ for $k\ne 1$, one needs to solve only Eq.~\ref{eq:dE2} and $\delta E_i/\delta \sigma^1=0$, an approximation which can be improved by inserting a small correction term.

\subsection{Transitions}

It is now possible to interpret the distinct phases introduced in Sec.~\ref{sec:latch} with the analytical tools just established.

The ability of the system to perform or not a transition is tightly linked with the possibility for some of the inactive units to flip in an active state and trigger a cascade, leading to an attractor transition.

\begin{figure}[ht]
\begin{center} 
\subfigure[]{\includegraphics[width=.23\textwidth]{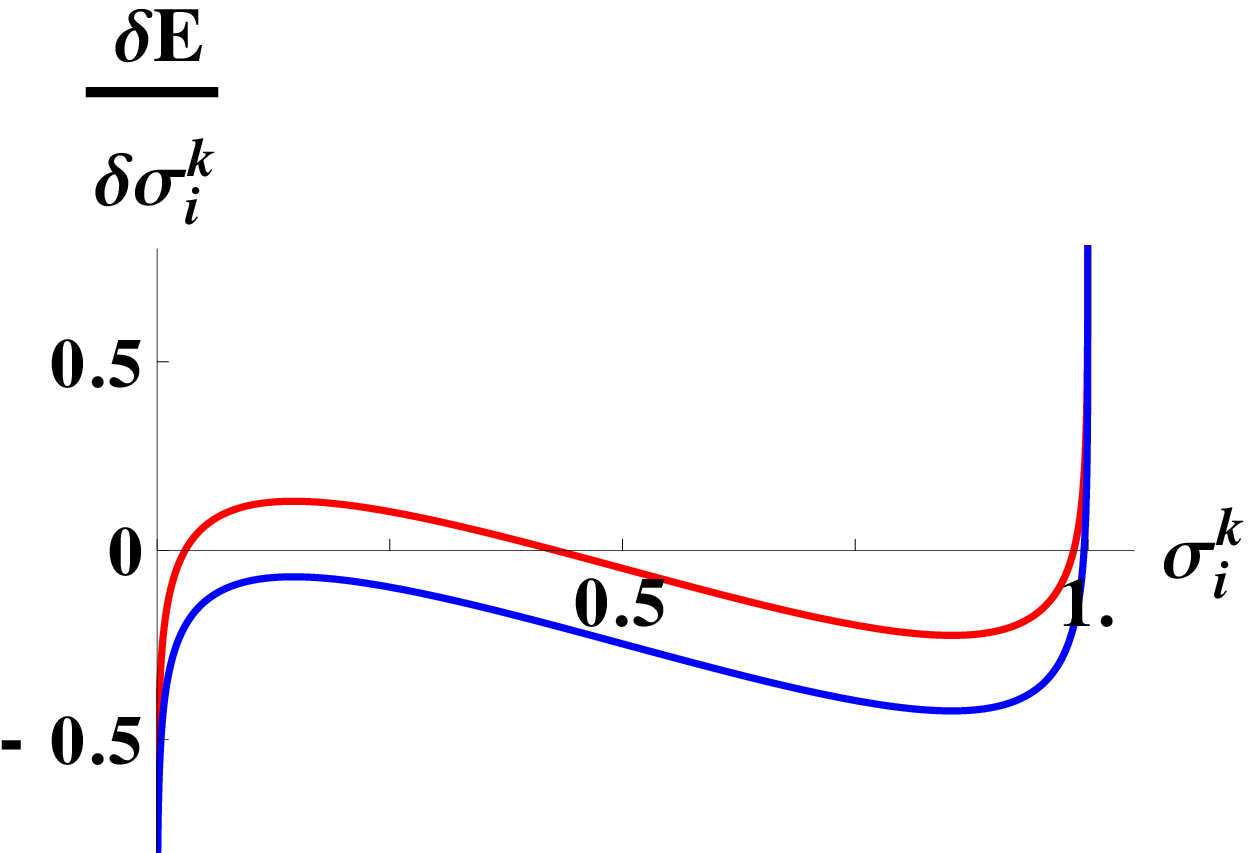}}
\subfigure[]{\includegraphics[width=.23\textwidth]{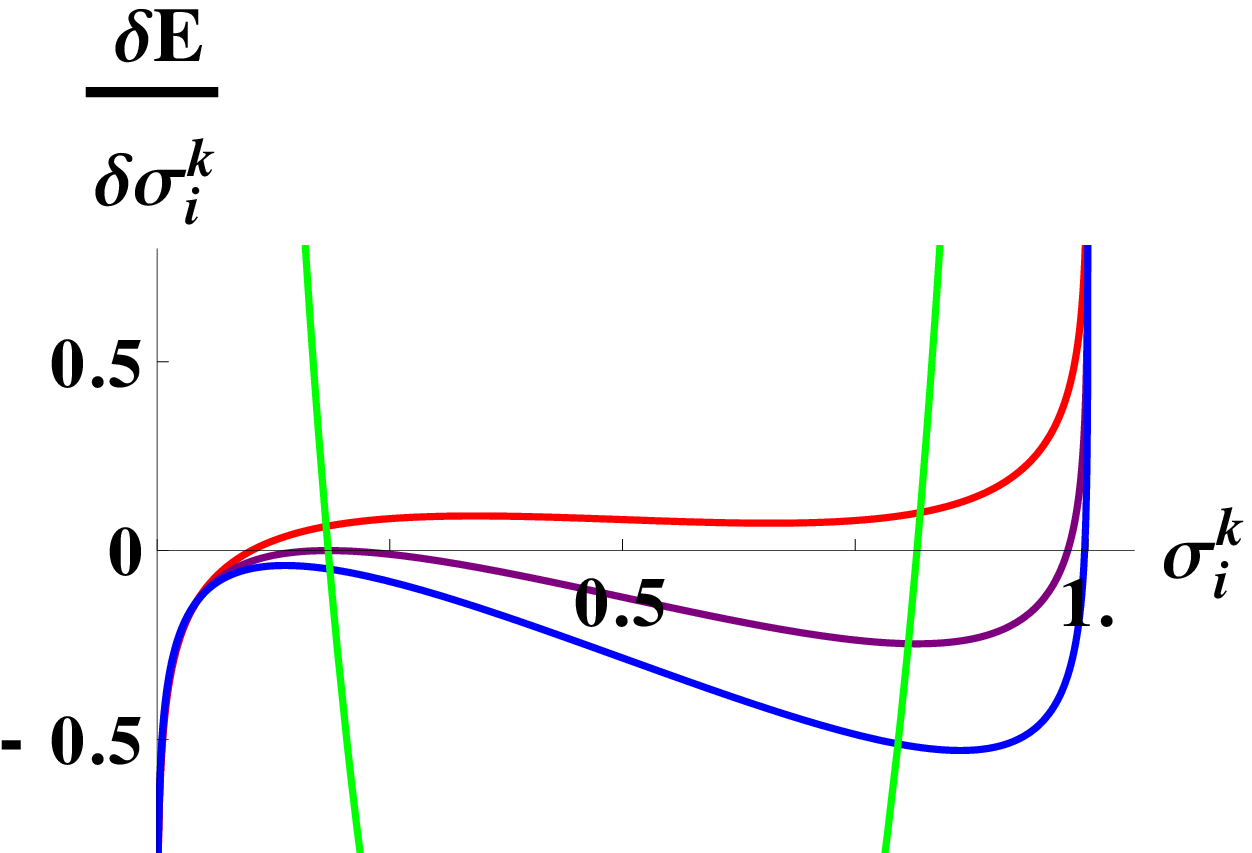}}
\caption{(Color online) Partial derivative of the Energy Functional with respect to $\sigma^k_i$. (a) Example of derivative which allows (lower, blue curve) or not (upper, red curve) latching transitions, e.g. because of different values of $\theta^0$; (b) derivative in the cases of $w=1.0$ (upper, red curve), $w=1.5$ (intermediate, purple curve) and $w=1.9$ (lower, blue curve), with the other parameters set at $S=9$, $N=600$, $p=140$, $C=90$, $U=0.1$, $a=0.25$ and $\beta=5$. The light green curve is the stability equation, Eq.~\ref{eq:d2Eidt}.
}\label{fig:blu_rosso_w}
\end{center} 
\end{figure}

In Fig.~\ref{fig:blu_rosso_w}(a) we plot Eq.~\ref{eq:dEidt} with respect to $\sigma^k_i$, for two different parameters sets (blue and red curves). In the case of the blue curve the energy derivative is negative until the unit reaches a complete activation. For the red curve, instead, it becomes positive already for small values of $\sigma^k_i$. The $i$ unit then, inactive in the retrieved attractor, will activate only if the local maximum of the energy derivative is under the $x$-axis, as in the blue case. This condition occurs only for some values of the parameters. Eq.~\ref{eq:dEidt} depends, indeed, explicitly on $w$, $S$, $U$, $T$ and implicitly, through $\sum_{j\ne i}^N\sum_{l=1}^S J_{ij}^{kl} \sigma^l_j$, on $a$, $C$, $p$, $N$. The variation of any of them moves and deforms the curve, producing the distinct phases presented in Sec.~\ref{sec:latch}.

Among all the parameters, we focus now on the $w$-$T$ dependence.

\subsection{The $w$-$T$ phase space}\label{sec:wt_ps}

Through computer simulations we explore the $w$-$T$ phase space. 
As shown in Fig.~\ref{fig:blu_rosso_w}(b), the qualitative effect of $w$ on the energy derivative is, if increased, to decrease the local maximum and, much more, the minimum of the function.

\begin{figure}[ht]
\begin{center} 
\includegraphics[width=.4\textwidth]{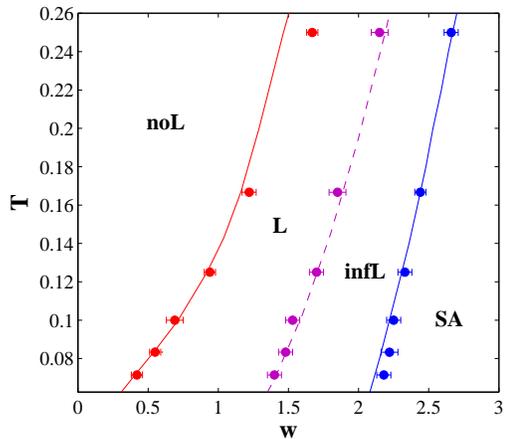}
\caption{(Color online) The $w$-$T$ phase space, in the slowly adapting regime. Dots refer to $w$ values, from simulations, that correspond to the beginning of the latching region (red, left), of the infinite latching region (purple, middle) and of the stable attractors region (blue, right). In these simulations $S=7$, $N=600$, $p=140$, $C=90$, $a=0.25$ and $U=0.1$. Solid curves refer to the analytical result (see Sec.~\ref{sec:curve}), while the dotted curve is just a guide for the eyes. Note that, in this regime, the $w$-interval between latching onset and the putative percolation transition, to latching sequences of duration proportional to $\tau_3$, is quite short, within the "error bars" of the simulation data.}\label{fig:sp}
\end{center} 
\end{figure}

Fig.~\ref{fig:sp} shows that, when changing the value of $w$, the system crosses several possible dynamics scenarios. Increasing $w$, the system is, respectively, able to retrieve a cued pattern but not to latch to other attractors ({\em no latching phase}); capable of a finite number of transitions, before getting trapped in the stable silent configuration ({\em finite latching phase}); capable of an infinite number of transitions ({\em infinite latching phase}); or trapped in the retrieved configuration ({\em stable attractor phase}). All this phase transitions correspond to critical conditions in the energy description.\\

\begin{figure}[ht]
\begin{center} 
\subfigure[]{\includegraphics[width=.23\textwidth]{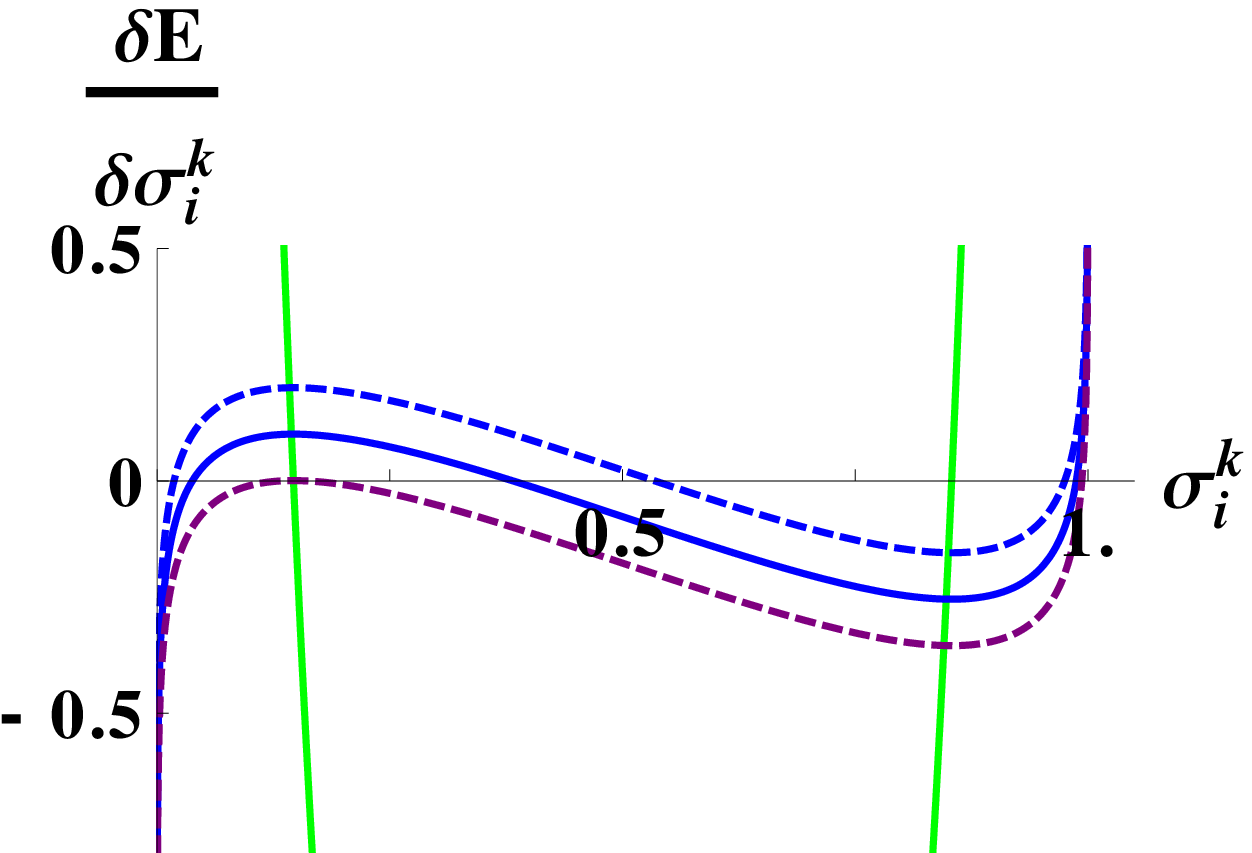}}
\subfigure[]{\includegraphics[width=.23\textwidth]{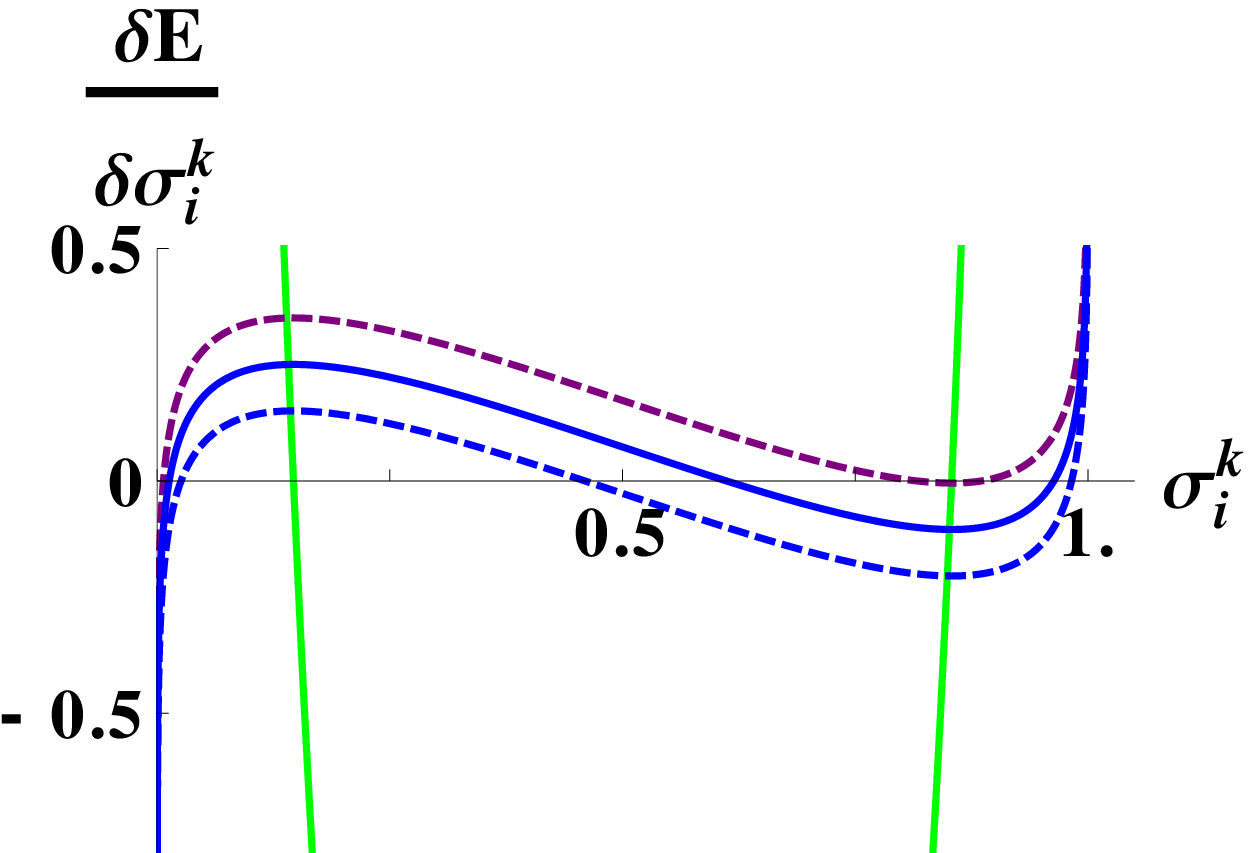}}
\caption{(Color online) Partial derivative of the Energy Functional, mean value (sold curve) $\pm$ standard deviation (dotted curve). (a) Critical condition in which some inactive units (lower, purple curve) are able to flip to an active state and guide a transition; (b) critical condition in which some of the active units (upper, purple curve) get deactivated, thus destabilizing the attractor.} \label{fig:ai}
\end{center} 
\end{figure}

{\em No Latching - Finite Latching transition.} It corresponds to the marginal condition in which the maximum is exactly on the $x$-axis (Fig.~\ref{fig:ai}(a)). For lower values of $w$, the derivative is positive and the system cannot latch; for higher values, the derivative is negative and the transition occurs. Whether the maximum is on the axis or not may depend of course on the initial condition, in particular on which pattern is retrieved by the external cue. If it is an early latcher, so to speak, the network may not continue to latch, or it may latch for only a few other steps. If instead $w$ is large enough to allow most attractors to latch, latching will typically last longer. Eventually, all attractors will lead to latching. 
If this is the case, transitions come one after the other, until the slow $\theta_i^0$, originally set at $0$, slowly increases (as we are focusing now on the slow adaptation regime in which $\tau_3 > \tau_2 \gg \tau_1$), bringing up the whole curve. 
As soon as the maximum reaches the $x$-axis, no unit is able to flip any longer, and transitions stop. The $w$-interval between early and late latchers is quite short, as seen by the standard deviation in the Fig.~\ref{fig:sp}.\\

{\em Finite - Infinite Latching transition.} Since the generic threshold $\theta_i^0$ can take as maximal value $\theta_i^0=1$, beyond a certain value of $w$ the growth of $\theta_i^0$ is not sufficient to ever make the maximum positive. For each value of the temperature $T$, then, there is a critical value $w$, beyond which latching transitions never stop.\\

{\em Infinite latching - Stable Attractors transition.} The increase of $w$, however, has the simultaneous effect of stabilizing network attractors. Therefore, increasing the self-reinforcement term the system reaches a condition in which, after the first retrieval, it remains trapped in the attractor. Analytically we can treat this last case similarly to the previous one, with the only difference that the energy derivative, formally identical to Eq.~\ref{eq:dEidt}, is now taken with respect to the activation value of an active unit. From Fig.~\ref{fig:ai}(b), it is possible to see that a generic active unit is stable until the Energy derivative has its minimum under the x-axis. Then, when the adaptive threshold $\theta_i^k$ increases, following the unit activation, it translates the curve upwards, inactivating the unit as soon as the local minimum crosses $0$. If $w$, however, is too large, the translation due to $\theta_i^k$ is not sufficient for the curve to cross the x-axis. There is then a critical value for $w$, beyond which all network attractors become stable (Fig.~\ref{fig:ai}(a)).

\subsection{Analytical Curves} \label{sec:curve}
Having set up an analytical framework, we can now try to explain the boundaries between the different dynamical phases obtained through computer simulations, and shown in Fig.~\ref{fig:sp}. 

As described in section \ref{sing_unit_term}, 
at the instability, Eqs.~\ref{eq:dEidt} and \ref{eq:dE2} should be solved simultaneously, to yield the value of $\sigma^k_i$ and a relation between the parameters, expressing e.g. $w$ as a function of the others. 

Setting all other parameters, and imposing conditions appropriate to the different phases, it is then possible, to solve the system numerically and find the critical values $w_c$ at the phase boundaries. Before analyzing individual phase transitions, we can make a general approximation, whether unit $i$ is active or inactive and $k$ its future or present activation state: we take $\sigma^k_i+\sigma^0_i\simeq 1$. The entropy term of Eq.~\ref{eq:dEidt} can then be approximated as
\begin{eqnarray}
T\left(\ln {\sigma^k_i \over \sigma^k_i+\sigma^0_i} -\sum_{l=1}^S \ln {\sigma^0_i \over \sigma^l_i+\sigma^0_i}\right)\simeq T \log \frac{\sigma^k_i}{1-\sigma^k_i}.
\end{eqnarray}

{\em No Latching - Finite Latching transition.}
The units that lead this transition are those inactive in the current attractor, that are activated in the next attractor. Among those units, the first to be activated are those that were inactive also before the current attractor, e.g. in the stable condition in which the system is prepared before the external cue that triggers the initial pattern retrieval, so that their $\theta^k$ and $\theta^0$ thresholds are still at, or close to, 0. In the unit-to-unit variability expressed by the fluctuations around the mean values in Fig.~\ref{fig:ai}, these units are at the lower end of the range: their activation then reduces the signal, i.e. lowers $\delta E_i/\delta \sigma^1$, for other units, which follow suit. To estimate where is the lower end of the fluctuation range, one may consider the variability expressed by Eq.~\ref{noise_amp} and the number of units to be activated, which are a subset of the $Na$ units active in the next attractor \textendash  those not active already. We find empirically that a good approximation to the $w_c(T)$ curve derived from the simulations is obtained by inserting a coefficient (possibly dependent on $N$) in front of the s.d. of the signal, and applying a correction to the fraction $a$ of units to be activated, dependent on $w$ and $T$, so that the mean signal is lowered by a term  
\begin{eqnarray}
1.7~\sqrt{\frac{(p-2)(a-\frac{wT}{2})}{C S^2}+\frac{\left(\bar{J\sigma}\right)^2}{C a}}
\end{eqnarray}
that we leave for future studies to fully justify. The numerical curve thus obtained is in reasonable agreement with that derived from simulations, as shown in Fig.~\ref{fig:sp}. Note that, except for higher $T$ values, the discrepancy between the curve thus obtained and simulation data is within the range of variability of the latter, i.e. within the difference between the very earliest latching, and the putative percolation transition.\\

{\em Finite - Infinite Latching transition.} 
The boundary between the finite and infinite latching regions is conceptually of the same nature, except that the phase transition occurs when $\delta E_i/\delta \sigma^1$ is lowered so much by the larger $w$ value as to compensate for the thresholds $\theta^0$ and $\theta^k$, which have grown during the latching sequence. Since their final values are difficult to determine, we do not attempt an analytical derivation of the phase boundary.\\

{\em Infinite latching - Stable Attractors transition.}
The transition from unstable to stable attractors is related to the inability of any active units to be deactivated by the increase in the adaptive threshold. If a few units go to the inactive state, others follow, with the same chain reaction process seen above. Therefore the condition for attractors to be stable is that no unit should be able to inactivate. The limit hence correspond to the condition in which, with maximal threshold, and $\theta^k=\theta^0=1$, the minimum of Eq.~\ref{eq:dEidt} lies on the x-axis. Again, one estimates the height of the $\delta E_i/\delta \sigma^1$ curve for the early deactivating units from the amplitude of the fluctuations in Eq.~\ref{noise_amp}. The result of this approximation is the blue curve in Fig.~\ref{fig:sp}. To produce this curve, we used directly mean $\theta$ values resulting from the simulations.

\subsection{Adapting the analysis to the fast adaptation regime}

The above analysis can be extended to the situation in which the specific thresholds $\theta^k_i$ continue to evolve on a much slower time scale than the activation values $\sigma^k_i$, but the generic thresholds $\theta^0_i$ evolve faster, and the limit can be considered in which they adapt almost instantaneously to the activation values. Strictly in that limit, it was shown earlier that one can consider the modified energy functional of Eq.~\ref{eq:Efast}, where the $\theta^0_i$ have been integrated out, and there is an additional term $(\sum_k \sigma^k_i)^2/2$ for each unit, to account for their rapid adaptation. A pre-factor $\mu < 1$ may also be inserted before the additional term, to take into effective consideration a non-instantaneous adaptation of the generic thresholds, i.e. when $\tau_3 < \tau_1$, but still the two time scales are comparable. It is simpler, though, to restrict ourselves to the limit $\tau_3 \ll \tau_1 \ll \tau_2$.

In this limit, the partial derivative of the energy functional with respect to any of its activation values remains numerically the same as in Eq.~\ref{eq:dEidt}, with the difference that the simple term $\theta^0_i$ now stands for $\sum_k \sigma^k_i$. As a consequence, the second derivatives that enter the stability matrix change, and they all acquire an additional term equal to 1. The components of the unstable eigenvector in the non-preferred directions are modified into
\begin{equation}
\epsilon^k \simeq \sigma^k \left( \beta w +\beta S +2S/\sigma^0 \right) + O(\bar{\sigma}^2);
\end{equation}
and, importantly, the stability condition now reads, to leading order,
\begin{equation}
\beta w \left(1 - {1\over S}\right)
\simeq \beta +{1\over \sigma^0} + {1\over\sigma^1} + O(\bar{\sigma}). \label{eq:dE2fast_ad}
\end{equation}
The additional $\beta $ term in Eq.~\ref{eq:dE2fast_ad} implies that the green curve in Fig.~\ref{fig:blu_rosso_w} is much higher, hence it takes a higher value of the local feedback coefficient $w$ to reach a phase boundary. The phase diagram of Fig.~\ref{fig:sp} is basically shifted to the right. Moreover, in the fast adaption regime there is essentially no finite latching region: it is squeezed into a tiny transition strip between the onset of latching, when starting in a favorable initial attractor, and the percolation-type transition to infinite latching, when latching proceeds from any attractor. These analytical expectations are well matched by simulation results (not shown), although a quantitative agreement, when e.g. $\tau_3 = 0.3 \cdot \tau_1$ does not really vanish, requires inserting a pre-factor to multiply the new term, as indicated above, e.g. $\mu \simeq 0.8$.

\subsection{Latching with correlated patterns}

\begin{figure}[ht]
\begin{center} 
\includegraphics[width=.4\textwidth]{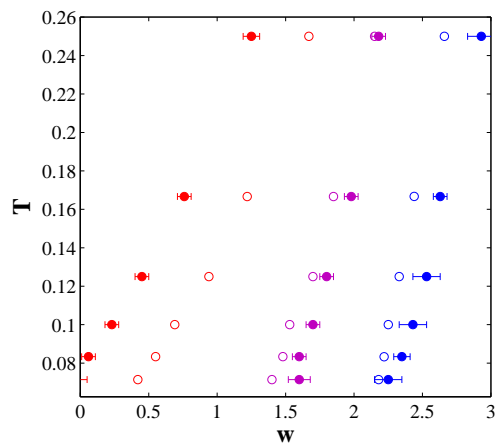}
\caption{(Color online) The $w$-$T$ phase space for a correlated set of patterns. Similarly to Fig.~\ref{fig:sp}, dots refer to $w_c$ values dividing the different regions of latching dynamics. Empty dots, the same as in Fig.~\ref{fig:sp}, refer to a randomly generated set of patterns, full dots to a weakly correlated set ($\zeta=10^{-6}, a_{pf}=0.4$).}\label{fig:spc}
\end{center} 
\end{figure}

Whereas to treat the fast adaptation regime one has to modify the single unit terms in Eq.~\ref{eq:dEidt}, to describe the phase diagram appropriate to the storage of correlated attractors one has to consider an enhanced variability, due to correlations, in the local field from the rest of the network, i.e. a larger standard deviation $\Delta_{J\sigma}$. Correlations produce more marked differences between "preferred" and "disfavoured" patterns, in their ability to lead to a latching transition; the distance between the dotted standard deviation curves in Fig.~\ref{fig:ai} is larger, and consequently latching tends to start off earlier, as a function of an increasing $w$, and to self-sustain, and eventually to be blocked by attractor stability, later. We do not attempt here an analytical estimate of the $\Delta_{J\sigma}$ produced by our algorithm for generating correlated patterns, but simulations show, in Fig.~\ref{fig:spc}, that the boundaries in phase space are modified as expected. The extent of the latching region is therefore amplified, already in the situation shown, with only weakly correlated patterns. With stronger correlation, the finite latching phase becomes even wider, but to consider the functionality of the associative network one has to remember that retrieval capacity can be seriously impaired, as indicated in Fig.~\ref{fig:storage}.

\section{Implications and further work}

We study here how the introduction of an adaptive element affects the dynamics of an autoassociative neural network. The model draws a sketch of the cortex as a network of Potts units, each of them representing a cortical patch, and interacting with each other through recurrent connections (Sec.~\ref{sec:pottsunits}). The stability of the stored memories, which are encoded in the connections between the units, is weakened by adaptive thresholds, which oppose the permanence in the current attractor. In this way, the network may be pushed out of the basin of attraction of the retrieved memory, and then it can slide towards a new one, performing what we have named a {\em latching transition} (Sec.~\ref{sec:Adaptivedynamics}).  

The introduction of adaptation is found to lead to a variety of phases of network dynamics. Depending on its parameters, the network is either unable to perform any transition, remaining inactive after the retrieval of a cued memory; or it does go through some transitions, jumping from memory to memory in a sequential retrieval chain that terminates in the inactive state; or through endless transitions, when the latching process self sustains indefinitely in time (Sec.~\ref{sec:latch}).  In particular, each of these phases can be reached by changing the number of local attractors assumed to exist in each cortical patch ($S$), the number of memories stored in the global network ($p$), the connectivity between cortical patches ($C$), the local feedback ($w$) or the noise internal to each patch ($T$) (Sec.~\ref{sec:phas-ton}).

The infinite latching region is however found not to be always accessible, and its extension in parameter phase space to be restricted by the presence of two other dynamics regions.
The increase in memory load of the network, which would favour a sustained latching chain, also impairs the ability of the system to retrieve individual memories. The so-called {\em storage capacity} (indeed, retrieval capacity) limit can be crossed either after the indefinite latching region is reached, and then it serves as its upper boundary, or before, and then it prevents the system from ever entering such a phase (Sec.~\ref{sec:sustained}). On the other hand, attractor stability can also be enhanced by those same manipulations that favour latching. As we saw in section \ref{sec:wt_ps}, an increase in the feedback parameter, $w$, leads to a condition in which attractors are so reinforced, as to become stable. This sets an alternative phase boundary to the latching region.

The introduction of adaptive elements brings a complication in the analytical description. The {\em effective field} imposed on a unit comes to depend on the time spent by the unit itself in that particular activation state (and in preceding states, as well). As a consequence, it is impossible to write down a function of state governing the dynamics of the system. To bypass this problem, we largely restrict ourselves to a slow adaptation regime and we treat separately the fast attractive from the slow adaptive dynamics (Sec.~\ref{sec:Adaptivedynamics}). The time scale separation allow us to focus on the specific instant when attractors begin to switch, and to derive analytically an approximate phase boundary between the no latching and latching conditions and the indefinite latching and stable attractor conditions, in the $w$-$T$ phase space (Sec.\ref{sec:curve}). The analytical treatment can also be adjusted to fit a type of fast adaptation regime, intended to model rapid local inhibition.

In this somewhat complex scenario, we have finally considered how correlations among patterns may affect network dynamics. With a two-step algorithm, we generate sets of patterns with different distributions of pairwise correlation (Sec.~\ref{sec:corrpatt}). Stronger correlation is found to decrease the storage capacity of the network, and to shrink their basins of attraction (Sec.~\ref{sec:storage}). On the other hand the very same destabilization brought about by correlation makes the system more prone to latching transitions. We sketch the $w$-$T$ phase space in the case of weakly correlated patterns, and we see that enhanced correlation widens the latching phase, by both lowering the value of $w$ at the no latching-latching boundary and increasing the value needed to stabilize attractors.

\subsection{Applications to analyses of cortical processing}

Despite its high level of abstraction, the model can serve as a mathematically well-defined benchmark, where to test general hypotheses about cortical dynamics. An obvious example is any relation hypothesised to exist between the various parameters mentioned above and the duration and characteristics of structured, spontaneous, non-stimulus-driven cortical activity.  Experimental paradigms have been used to explore such activity, even in non-human primates, e.g. in the free drawing task developed by Moshe Abeles and coworkers \cite{Shm+05}. The behavioural model and the mathematical model can be brought to bear onto each other. Further work is needed, however, to understand how to experimentally manipulate and control any of the parameters, in particular the non-structural ones, including the storage load $p$, or the local noise level $T$, or the relative degree of local feedback $w$, so as to investigate any change induced in the dynamics. 

Another issue that can be probed with the model is the relation between the notions of a {\em global workspace} \cite{Deh+98} and of serial {\em vs.} parallel processing \cite{Del+07}. At the qualitative level of information processing models, the availability of a global workspace can be taken to imply, irrespective of the specialization of individual components \cite{Kou+09} that one and only one information processing operation can occur within it at any given time. Hence, a single interconnected network may appear to force serial processing: the lack of segregating boundaries, instead of being a resource, is turned into a nuisance; it imposes a bottleneck. Items that have been processed in parallel along separate sensory pathways can, on their way to distinct motor pathways, get stuck, so to speak, through a router, i.e. a distinct piece of cortical machinery with {\em ad hoc} properties \cite{Zyl+10}, which, like any centralised bureaucrat, acts as a bottleneck \cite{Pal+11}.

The Potts model, in contrast, with its generic autoassociative architecture, indicates that lack of anatomical segregation is not necessarily equivalent to serial processing. Processing is effectively serial, in the model, when latching dynamics is particularly clean, involving a single global attractor at a time, a situation which does occur in certain regions of parameter space. In other regions, however, dynamics is messier, with several attractors simultaneously active, and latching transitions that involve only part of a structurally complex, but non-segregated network. A rapid transition from attractor A to D may occur while attractors B and C are also active, over units largely distinct from those active in either A or D. Can the system in this case be argued to operate along three separate streams? Hardly so, because the streams are not segregated, and their support basis, i.e. the corresponding subset of active Potts units, changes all the time. The distinction between serial and parallel therefore ceases to be as clear cut as it normally is in engineering systems, and it becomes a {\em quantitative} rather than qualitative issue. The whole network can be regarded as a single interactive workspace for multifarious parallel processes, like a modern urban society. These aspects require a dedicated study, that adapts information theoretical measures to assess the dependence between successive attractor states \cite{Rus+11}.

A third domain of applicability of the model is in studying the relation between local and global attractors. In the Potts model, local attractor states are incorporated in the definition itself, which is regarded as a reduced or {\em effective theory} for a full-fledged neuronal model \cite{Akr+11}. The search for experimental evidence for such states has had clearer successes in the hippocampus \cite{Wil+05,Jez+11}, than in higher visual areas of the cortex \cite{Ami+97,Akr+09}. Nevertheless, data compatible with attractor-like behavior is appearing in a variety of preparations, analysed with different approaches, from Hidden Markov Models applied to monkey frontal cortex \cite{Abe+95} and to the rat gustatory system \cite{Jon+07} to kernel methods applied to multi-unit activity from rat anterior cingulate cortex \cite{Bal+11}. A recurrent question is whether such signatures reflect sequences of global attractors states, as have been hypothesized to underly also perceptual decisions in humans, as discrete {\em moments of thought} \cite{Gra+11}. A full mechanistic understanding of such phenomenology seems to require multi-level neuronal network models, where both local and global scales (and possibly intermediate ones) are represented explicitly. The complexity of such model, however, is daunting, so that the Potts model can provide a more tractable approach.

\subsection{Evolutionary perspective}

An evolutionary perspective \cite{Kru+05} can make use of the Potts model to explore the hypothesis \cite{Tre05} that a critical step leading to human forms of cognition is a phase transition, incurred as a result of a quantitative change in some of the parameters characterizing the primate (or indeed mammalian) cortex. In this hypothesis, no major structural change differentiates human cortical networks from those of at least fellow great apes, but a qualitatively different functionality of the very same networks may have arisen out of an increase in size, or in connectivity, that has induced a phase transition. The Potts model offers several candidate transitions: from no latching to finite latching, from finite to infinite latching, as well as the percolation-type of transition that, in the slowly adapting regime, follows closely the onset of latching. Identifying the most appropriate scenario is made somewhat complicated by the fact that a plausible cortical model would presumably include both rapidly adapting and slowly adapting thresholds \cite{Akr+11}, and the two opposite limit cases would have, at the very least, to be combined. This is left for further work. Nevertheless, the notion that spontaneous cortical activity can proceed longer in a larger or more connected brain (not in a scaling relation, but dramatically longer, as after a phase transition) resonates with the notion of infinite recursivity, put forward in linguistics \cite{Cho+02} and in other domains \cite{Jac09,Hes+04,Ama+07}, as we have discussed elsewhere \cite{Rus+11b}.

In this view, the parameters that may constrain cortical functionality are those that are kept at relatively low values, in non-human species, for structural reasons. Indeed, some of the parameters in the Potts model can be thought to reflect quantities that can be freely set at optimal values, and these may include the relative local feedback strength $w$, the effective noise level $T$, or the adaptation time scales \cite{Mar10}. Other parameters in the model, including the number of patches $N$, their connectivity $C$, the number of attractor states in each patch, $S$, or the storage load $p$, reflect cortical quantities that it may be more difficult to act upon, either because increasing their values requires reformulating or extending developmental programmes, or because of their mutual relationships. Thus, for example, increasing the memory storage load is always possible, but it is of little use if, as a result, the system becomes overloaded, beyond its storage capacity, set mainly by the connectivity and by the density of local attractor states. Therefore an increase in $p$ may only be enabled by a concurrent increase in $C$ and/or $S$ \cite{Rus+11b}. 

It would be desirable, then, to consider the phase-transition-to-human-cognition hypothesis by studying the phase diagram of the Potts model in relation to the parameters $C, N, S, p$. Unlike $w$ and $T$, which affect mainly the single-unit terms, these parameters enter largely (except for a minor feedback role for $S$) in the signal conveyed by the rest of the network, and in particular they determine its variability $\Delta_{J\sigma}$. The functional dependence of $\Delta_{J\sigma}$ on $C, N, S$ and $p$ is however quite different in different models of the pattern generation process. It can be estimated analytically for randomly correlated patterns, and numerically for patterns generated by our multi-factorial process, but other models for this key process may be proposed that would lead to rather different effects on the variability of the signal arriving at individual cortical patches, and hence on the phase boundaries of the system. This challenging issue is left for future investigations.

\begin{acknowledgments} 
We are grateful to Yasser Roudi, Emilio Kropff, Hossein Abbasian and Mohammed F Abdollah-nia, Mohammed Karim Saeed-Ghalati (also for sharing with us their preprint {\em Optimal region of latching activity in an adaptive Potts model for networks of neurons}) and Andrea Gambassi for several useful discussions and encouragement.
\end{acknowledgments}

\bibliography{Marzox}
\bibliographystyle{apsrev}







\end{document}